\documentclass[useAMS,usenatbib]{mn2e}
\usepackage{subfigure}
\usepackage{natbib}
\usepackage{graphicx}
\usepackage{lscape}
\usepackage{rotating}
\usepackage{times}
\usepackage{epsf}
\usepackage{fancyhdr}
\usepackage{longtable,lscape}
\usepackage{amssymb,amsmath}
\usepackage{color}

\newcommand{\xmmn}{{\it XMM-Newton~\/}}

\newcommand{\ergscm}{erg\,cm$^{-2}$\,s$^{-1}$}

\newcommand{\ergs}{erg s$^{-1}$}

\newcommand{\Lx}{$L_{\rm X}$~}

\def\eg{{e.g.~\/}}

\def\ie{{i.e.~\/}}

\def\deg{{${ }^{\circ}$}}

\title[Low-\Lx sources and the GRXE]{Low-luminosity X-ray sources
and the Galactic ridge X-ray emission}
\author[R.~S. Warwick]{R.~S. Warwick$^{1}$\thanks{E-mail:rsw@le.ac.uk} \\ 
$^{1}$Department of Physics and Astronomy, University of Leicester, 
University Road, Leicester, LE1 7RH, UK}

\begin{document}

\date{Accepted. Received; in original form}

\pagerange{\pageref{}--\pageref{}} \pubyear{2014}

\maketitle

\label{firstpage}

\begin{abstract}

{\color{black} Using the \xmmn Slew Survey, we construct a hard-band
selected sample of low-luminosity Galactic X-ray sources.
Two source populations are represented, namely coronally-active stars and
binaries (ASBs) and cataclysmic variables (CVs), with
X-ray luminosities collectively spanning the range $10^{28-34}$ \ergs (2--10 keV).
We derive the 2--10 keV X-ray luminosity function (XLF) and volume emissivity
of each population. Scaled to the local
stellar mass density, the latter is found to be 
$1.08 \pm 0.16 \times 10^{28}  \rm~erg~s^{-1}~M_{\odot}^{-1}$ and
$2.5 \pm 0.6 \times 10^{27} \rm~erg~s^{-1}~M_{\odot}^{-1}$, 
for the ASBs and CVs respectively,
which in total is a factor 2 higher than previous estimates.
We employ the new XLFs to predict the X-ray source counts on the Galactic plane at
$ l = 28.5^{\circ}$ and show that the result is consistent with current
observational constraints. The X-ray emission of faint, unresolved ASBs
and CVs can account for a substantial fraction of the Galactic
ridge X-ray emission (GRXE). We discuss a model in which $\sim80$ per cent of
the 6--10 keV GRXE intensity is produced in this way, with the remainder
attributable to X-ray scattering in the interstellar medium and/or 
young Galactic source populations. Much of the hard X-ray emission attributed
to the ASBs is likely to be produced during flaring episodes.}

\end{abstract}


\section{Introduction}
\label{sec_1}

The hard Galactic ridge X-ray emission (GRXE) can be traced from
the Galactic Centre, where its surface brightness peaks, along the 
Galactic plane out to $ |l| = 60^{\circ}$ and beyond
(\citealt{wor82}; \citealt{war85}; \citealt{koy86a}; \citealt{yam93};
\citealt{yam96}; \citealt{rev06}; \citealt{koy07}). 
In the 4--10 keV band the spectrum of the GRXE resembles that of
5--10 keV optically-thin thermal plasma in collisional ionization equilibrium, 
with prominent lines arising from K-shell emission in He-like and
H-like Fe at 6.67 keV and 6.96 keV, respectively (\citealt{koy86a}; \citealt{kan97}).
An Fe K$\alpha$ line at 6.4 keV resulting from the fluorescence of
neutral or near-neutral Fe is also seen
(\citealt{koy96}; \citealt{ebi08}; \citealt{yam09}). 
Below 4 keV, emission lines of abundant elements
such as Mg, Si, S, Ar and Ca are evident, with the He-like to H-like
intensity ratios (taken together with those of Fe)
indicative of a multi-temperature plasma (\citealt{kan97}; 
\citealt{tanaka02}). The temperature structure appears
to be similar at different locations along the GRXE, with
the exception of the region within a degree or so of the
Galactic Centre \citep{uch11,uch13}. Above 10 keV the GRXE spectrum
may exhibit a hard powerlaw tail, possibly extending to the hard 
X-ray/$\gamma$-ray region (\citealt{yam97}; \citealt{val98};
\citealt{strong05}; \citealt{kriv07}).

The GRXE has been interpreted both in terms
of the superposition of faint point sources (\citealt{sug01};
\citealt{rev06}; \citealt{yuasa12}) and as a highly energetic, 
very high temperature phase of the interstellar medium (\citealt{koy86b}; 
\citealt{kan97}; \citealt{tanaka02}), amongst other possibilities 
(\eg \citealt{val00}; \citealt{mol14}). Recent studies have shown
that the GRXE surface brightness follows that of the near infra-red (NIR)
light associated with the old stellar population of the
Galaxy (\citealt{rev06}).  Also very deep {\it Chandra} observations
have directly resolved over 80\% of the GRXE near the Galactic
Centre into point sources (\citealt{rev09}; {\color{black} \citealt{hong12}}). This
is compelling evidence that the bulk of GRXE originates in the
integrated emission of point sources, although there is still some
debate as to whether there might remain some excess emission
attributable to a distinct very hot diffuse component within
a degree or so of the Galactic Centre (\citealt{uch11, uch13}; 
\citealt{heard13}; \citealt{nish13}). It has also been suggested recently
that some fraction the GRXE might arise from the scattering of the
radiation from bright Galactic X-ray binaries (XRBs) by the interstellar
medium (\citealt{mol14}).  

One requirement on any Galactic source population (or populations) deemed
responsible for GRXE is that the inferred volume emissivity should be
sufficient to give rise to the observed surface brightness of the GRXE.
A second constraint is that the integrated spectrum of the sources
should match the observed GRXE spectrum. In this context it has been
proposed that a mix of magnetic cataclysmic  variables (CVs) plus 
coronally-active binaries may have sufficient spatial density and
hard X-ray luminosity to account for the GRXE
(\citealt{muno04}; \citealt{saz06}; \citealt{rev06}; \citealt{rev08}). 
Also, CVs and active binaries have marked spectral similarities
to the GRXE (\citealt{rev06}; \citealt{tanaka10}; \citealt{yuasa12}).
However current uncertainties relating to the population
properties and statistics leave many of the details of this model
to be confirmed.  

One of the key elements in any source model of the GRXE is the
underlying X-ray luminosity function (XLF) of the contributing
sources, especially in the intermediate-to-low luminosity range
(nominally $10^{28-34}$ \ergs). To date the best determination
of the 2--10 keV XLF in this regime is that reported
by \cite{saz06}, based on the combination of {\it RXTE} and
{\it Rosat} survey measurements.  However, an obvious limitation
of this work, which is particularly important for the 
coronally-active systems, was the need to extrapolate from the soft 
(0.1--2.4 keV) bandpass of {\it Rosat} into the hard 2--10 keV band.
In this paper we remedy this situation by presenting
a new determination of the 2--10 keV XLF based on a hard-band selected
source sample derived from the \xmmn Slew survey (XSS; \citealt{sax08}). 

The remaining sections of this paper are organised as follows. In the 
next section we discuss the properties of a sample of
low-luminosity Galactic X-ray sources detected in the XSS in the
hard X-ray band.  This sample comprises coronally-active stellar sources
plus a smaller number of CVs (magnetic and non-magnetic).
In \S\ref{sec_3} we use this XSS-derived source sample to determine
a new estimate of the XLF of Galactic sources with 2--10 keV X-ray
luminosities in the range $10^{28-34}$ \ergs.
We go on (\S\ref{sec_4}) to predict the form the Galactic
X-ray source counts at Galactic longitude $l = 28.5^{\circ}$ and compare
the result with the available observational constraints.
We also estimate the integrated X-ray emission due to faint Galactic
X-ray sources and compare the result with the measured GRXE intensity
(again at $l = 28.5^{\circ}$). Finally, we discuss various
aspects of our results in the context of the origin of the GRXE (\S\ref{sec_6})
and then, summarize our conclusions (\S\ref{sec_7}).


\section{The source sample derived from the XSS}
\label{sec_2}

\xmmn  slews between successive pointings at a rate of 90$^{\circ}$ per hour.
This results in an exposure time for X-ray sources lying on the slew path
of typically between 1--10s.  As reported by \citet{sax08}, based on such slew
observations, the on-going XSS provides coverage of 
a significant fraction of the sky in a broad X-ray passpass.

In the XSS data reduction, the source count rate measured 
in the EPIC pn camera through the medium filter is recorded in
two nominal energy ranges, namely a soft (0.2--2 keV) band and a  
hard (2--12 keV) band.  The limiting sensitivity for source detection
in the hard band is approximately $3 \times 10^{-12}$ \ergscm~
(2--10 keV). This is roughly an order of magnitude deeper than the 
`classical' all-sky surveys  which are currently available in
this energy range, such as those from {\it Uhuru}, {\it Ariel V}, and
{\it HEAO-1} (\citealt{forman78}; \citealt{war81}; \citealt{mchardy81}; 
\citealt{pic82}; \citealt{wood84}). Further details of the XSS can be found
in \citet{sax08}.

In a recent paper, \citet{war12} analysed a sample of 487 sources detected in
the XSS hard band within a survey region encompassing $35\%$ of the sky
at high Galactic latitude ($|b| > 10$\deg).  Through 
cross-correlation with multi-waveband catalogues,  47\% of the
XSS sources were found to have likely
identifications with extragalactic sources (normal galaxies, active galaxies
and clusters of galaxies) with a further 16\% similarly identified
with stellar sources (coronally-active stars and 
accreting binaries). \citet{war12} further concluded that many (perhaps the
majority) of the remaining XSS sources with no obvious
identification were likely to be spurious (a natural consequence of the
low-significance threshold used to define the input sample).

\begin{figure}
\centering
\includegraphics[width=5cm,angle=-90]{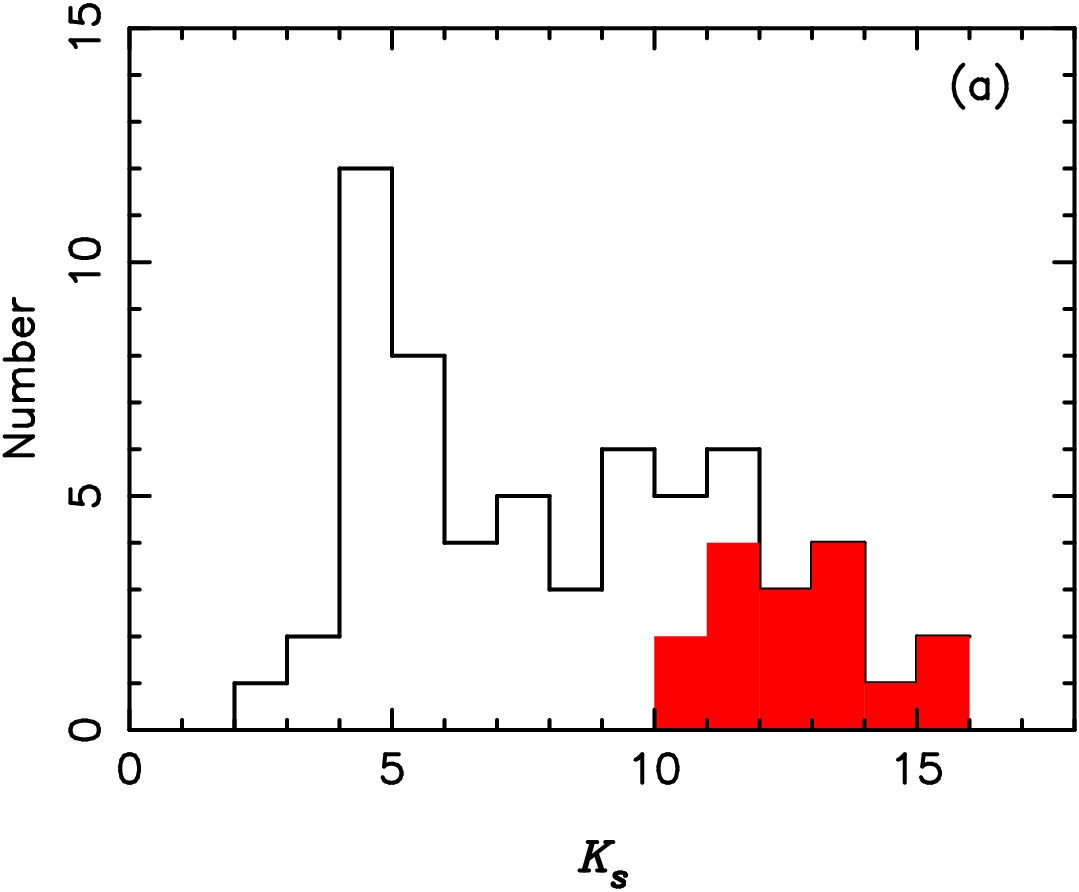}
\includegraphics[width=5cm,angle=-90]{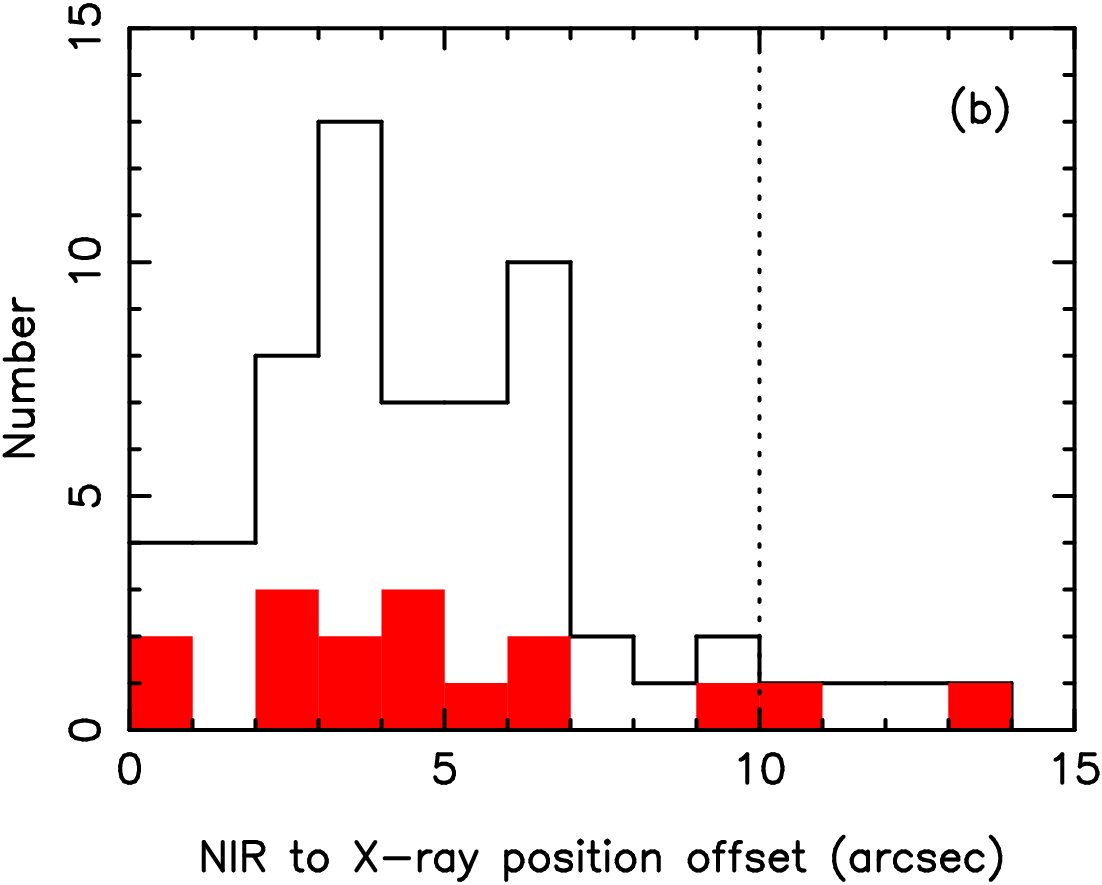}
\linespread{1}
\caption{(a) The $K_{s}$ magnitude distribution of the NIR counterparts to the
sources in the current sample; the region shaded red corresponds to the CVs,
with the remainder representing the ASBs.  (b) The spatial
offset between the 2MASS counterpart and the X-ray position; the red 
shading again corresponds to the CVs. The dotted vertical line is
drawn at 10 arcsec and represents a nominal 90\% X-ray error circle radius
for the XSS (\citealt{war12}).}
\label{fig_1}
\end{figure}

Making use of the facilities afforded by 
Vizier\footnote{http://vizier.u-strasbg.fr/viz-bin/VizieR/},
and Simbad\footnote{http://simbad.u-strasbg.fr/simbad/},
we have re-examined the 80 sources classed as ``Stars'' or ``Other'' in
the XSS sample considered by \citet{war12}. After excluding 11 sources
associated with LMXB/HMXB in Local Group galaxies, 4 supernova remnants, an
O star in Orion and 2 sources with no clearcut optical/NIR
counterpart, we are left with a sample of 62 hard-band selected X-ray sources
of likely Galactic origin.  This set of objects (hereafter referred to as the
{\it current} sample) includes 46 sources associated with
coronally-active stars and binaries (ASBs) plus 16 probable CVs.
All of these sources have a likely 2MASS\footnote{The Two Micron All 
Sky Survey (2MASS)  (\citealt{cutri03}; \citealt{skrutskie06})
provides uniform coverage of the entire sky in three NIR bands,
namely $J$ (1.25$\mu$), $H$ (1.65$\mu$) and $K_{s}$ (2.17$\mu$).} 
counterpart within or in close proximity to the X-ray error circle. 
Fig. \ref{fig_1} shows the distribution of the 2MASS $K_{s}$ magnitude for 
the current sample and also the distribution of the NIR to
X-ray position offsets.


\begin{figure}
\centering
\includegraphics[width=6cm,angle=-90]{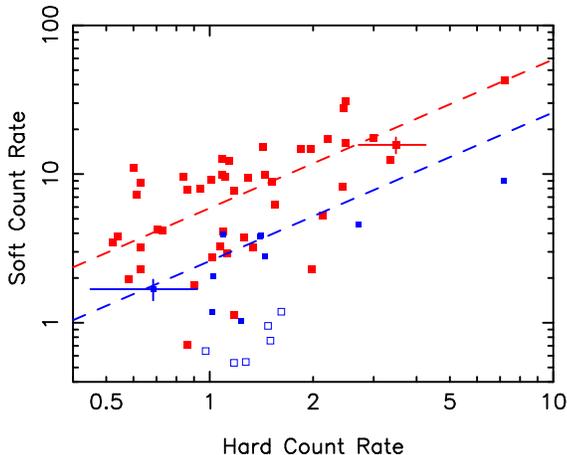}
\linespread{1}
\caption{{\color{black}The XSS soft-band count rate plotted versus the
XSS hard-band count rate for the ASBs (red points) and CVs (blue points) 
comprising the current sample. Upper-limit soft-band rates are identified
by open symbols.  Representative error bars are shown
for two of the sources. The dashed lines illustrate the ratio
of the hard-to-soft count rates predicted on the basis of the assumed
ASB and CV spectral forms (see text).}}
\label{fig_2}
\end{figure}


\subsection{The coronally-active stars and binaries}
\label{sec_21} 

Table \ref{tab_1} summarises some of the properties of the 46 ASBs in the
current sample. The columns of the table 
give the following information: the XSS catalogue name; the assumed distance; 
a reference to the distance estimate;
the type of coronally-active system (if known); the name of the star or 
binary system; the spectral type of the star(s) comprising
the system; the inferred (composite) absolute $K_{s}$-band magnitude ($M_{K}$);
an indication of whether the system includes either a giant or sub-giant star;
the measured X-ray count rate in the XSS hard band and, finally, 
the inferred 2--10 keV  X-ray luminosity (log $L_{\rm X}$).

The $M_K$ value is derived from the assumed distance and the 2MASS $K_{s}$ 
measurement {\it without} application of any reddening correction 
(as the impact of reddening will be small for the great majority
of the sources in this high-latitude sample). Similarly, the X-ray
luminosity is determined from the quoted distance and the XSS
hard-band count-rate based on the calibration,
$1~\rm ct~s^{-1} = 6.2 \times 10^{-12} \rm~erg~s^{-1}~cm^{-2}$ 
(2--10 keV). Here we have assumed the source spectrum can be represented
as a $T = 35$ MK ($kT = 3$ keV) {\bf apec} plasma with a
metal abundance $Z$ = 0.4 Z$_{\odot}$, subject to negligible line-of-sight
absorption. {\color{black} This spectral form is reasonably compatible
with the typical hard-to-soft count-rate ratio  measured in the XSS
for the ASB sample - as illustrated in Fig. \ref{fig_2} -  although
the scatter in this ratio is large.
The underlying spectral characteristics of ASBs are discussed further
in \S\ref{sec_6}.}

\begin{figure}
\centering
\includegraphics[width=6.7cm,angle=-90]{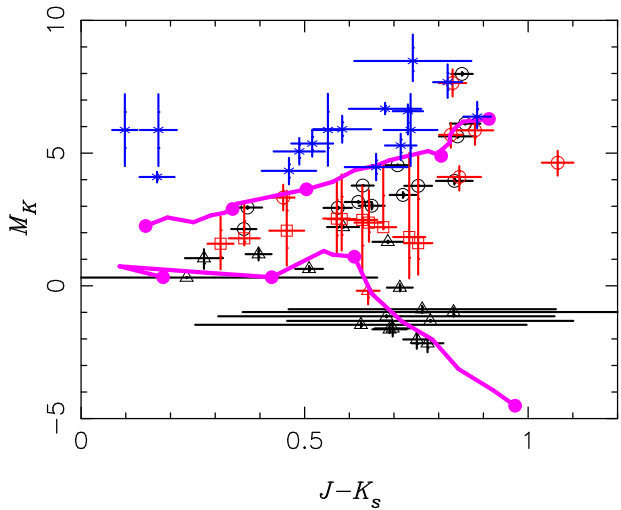}
\linespread{1}
\caption{The absolute $M_K$ magnitude versus the  $J - K_{s}$ colour
for the NIR counterparts of the sources in the current sample. The ASBs
with Hipparcos parallax measurements are shown as black symbols, with
systems known to contain a giant (or sub-giant) star represented as triangles 
and the rest as circles. The ASBs without parallax measurements are shown as
red symbols -- known giants
as triangles, known dwarf systems as circles and sources of 
uncertain type as squares. For the latter, the vertical red error
bars encompass the uncertainty in the dwarf versus giant classification.
The upper solid line corresponds to the locus of dwarf stars 
of spectral type F0 through to M5 (with the solid circles 
marking, from left to right, spectral types FO, GO, K0, M0 and M5). The lower
solid line similarly corresponds to the locus of giant stars of spectral 
type F0 to M0 (with the solid circles marking, from left to right, spectral
types F0, G0, K0 and M0). The blue asterisks symbols, plus associated
error bars, represent the CVs in the current sample.}
\label{fig_3}
\end{figure}

For the present study, the estimation of the source distance is key.
Of the sources in Table \ref{tab_1}, 29 have parallax measurements
from Hipparcos (\citealt{perryman97}; \citealt{van07a}; \citealt{van07b})
and hence have a reasonably precise and unambigous distance determination. 
Using the Hipparcos
distance together with the $J$ and $K_{s}$ magnitudes from 2MASS (again
without applying any correction for reddening), we can locate these 29 systems
in the $J-K_{s}$ colour versus absolute $M_K$ magnitude plane, as illustrated
in Fig. \ref{fig_3}. Note that although the 2MASS photometry errors 
typically result in an error on the $J-K_s$ colour of less than
$\sim 0.05$, for 6 very bright stars (with $K_s < 4.5$) saturation effects 
have presumably given rise to substantially higher uncertainties.  
Fig. \ref{fig_3} also shows the loci of dwarf and giant stars
as derived from the stellar spectral-flux library compiled by \citet{pic98}
(see their Tables 2 and 5), after applying the colour transformations
(Bessell \& Brett system to 2MASS) reported by \citet{car01}.
Evidently virtually all of the ASBs with Hipparcos
distances lie within the band defined by these two loci (the one
exception being the M3.5V star associated with XMMSL1 J184949.8-235011,
which also happens to be the nearest star in the sample). 

Based on information in the literature together with
the source location in Fig. \ref{fig_3}, we find that 16 out of the 29
coronally-active systems with Hipparcos distances most likely contain 
a giant or sub-giant star. Within the subset of 29
sources, there are 22 confirmed or probable binaries, comprising either
RS CVn, BY Dra, W UMa or Algol type systems (\citealt{gudel04}).

The 17 ASBs lacking Hipparcos parallaxes, as might
be expected, tend to be the sources at fainter end
of the $K_S$ magnitude distribution. 
Using information in the literature we were able to classify 
6 of these systems as likely containing late-K or M-dwarf stars\footnote{
One of this group, the counterpart to XMMSL1 J160900.7-190846, may in
fact be a pre-main sequence late-K dwarf star at the near-edge of
the Upper Scorpius OB Association (\citealt{car09}); on the basis of
its extreme $J-K_{s}$ colour this star appears to be 
significantly reddened.}.  A further source is a likely 
RS CVn binary containing a G7III star (\citealt{kir12}).
For these 7 systems we have either taken the distance estimate directly
from the literature or estimated $M_{K}$ from either the dwarf
or giant locus  in  Fig. \ref{fig_3} (taking due note of 
the measured $J-K_s$ colour). In the latter case
the distance was then determined via the 2MASS $K_{s}$ magnitude and 
a distance error range assigned assuming an uncertainty in $M_{K}$ of
0.5 magnitude.

For the remaining 10 sources lacking parallax measurements
the information in the literature relating to the spectral type,
luminosity class and/or distance was very limited. Without any clear
indication as to whether or not a given system contains a giant
or sub-giant star, 
we set its $M_{K}$ at the mid-value between the dwarf and
giant loci in  Fig. \ref{fig_3} (for the given  
$J-K_{s}$ colour), with a preliminary error range in  $M_{K}$
equal to the separation of the two loci.
The corresponding distance and  the distance error range were then
calculated from the observed $K_{s}$ magnitude.

\begin{figure}
\centering
\includegraphics[width=6cm,angle=-90]{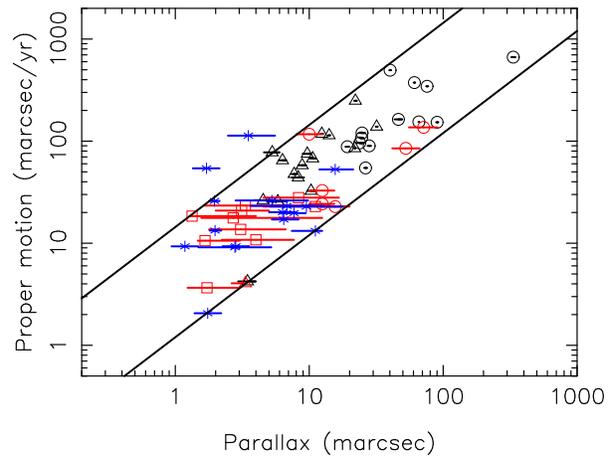}
\linespread{1}
\caption{The proper motion in marcsec yr$^{-1}$ plotted against the
parallax in marcsec (measured or inferred) for the NIR counterparts of
the sources in the current sample. 
The symbols are the same as in Fig. \ref{fig_3}.
The two diagonal lines represent nominal upper and lower bounds to the
scatter and correspond to transverse velocities of 68 and 
5.7 $\rm km~s^{-1}$, respectively.}
\label{fig_4}
\end{figure}

We have also compared the measured proper motion  (largely taken
from the PPMXL catalogue; \citealt{roeser10})  with the parallax inferred
from the assigned distance - see Fig. \ref{fig_4}.
Using the 29 sources with Hipparcos measurements as a control sample
to define the upper and lower bounds to the scatter in the
proportionality between the two axes in Fig. \ref{fig_4}, we were, at least
in a few cases, able to set an additional constraint on the source distance.
As a final step we also applied a limit to the distance upper-bound by
restricting the inferred 2--10 keV X-ray luminosity for these coronal emitters
to less than 10$^{32.5}$ \ergs.

Table \ref{tab_1} reports the estimated distance and distance error
range for each of the 17 ASBs lacking parallax measurements. The position of
these 17 sources in the NIR colour-magnitude diagram is shown in
Fig. \ref{fig_3}, together with the commensurate final error bars on $M_{K}$.

\begin{figure}
\centering
\includegraphics[width=8cm,angle=-90]{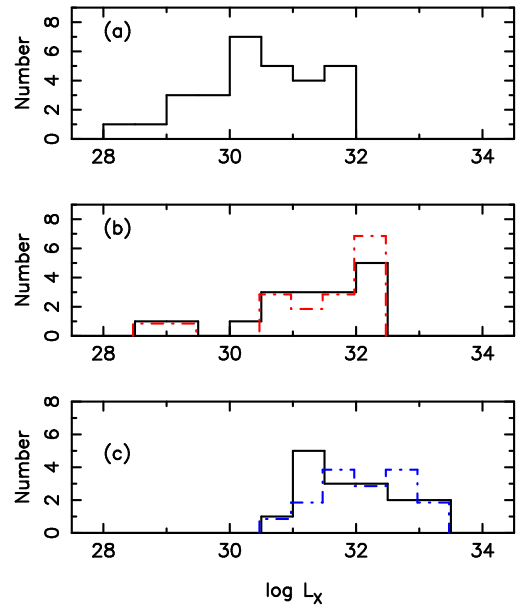}
\linespread{1}
\caption{{\it Panel (a):} The distribution of X-ray luminosity for
the ASBs with Hipparcos parallax measurements; {\it (b):} The same plot for
the ASBs without Hipparcos parallax measurements. The dashed (red) histogram
shows the impact of revising the source distance to the upper bound
of its range in the case of the 10 sources for which the dwarf-to-giant
classification is uncertain. {\it Panel (c):} The distribution of
X-ray luminosity for the CVs.  The dashed (blue) histogram
shows the impact of revising all the CV distances to their
upper bound.}
\label{fig_5}
\end{figure}

As noted above, of the 29 systems with parallax measurements, 16, \ie 55\%, 
contain a giant or sub-giant star. In contrast the statistics for the
systems without parallax measurements are 6 dwarfs, 1 (likely) giant
and 10 of uncertain type. On the basis of this comparison, one might
conclude that the majority of the latter are likely to be systems
containing giant (or sub-giant) stars. Hence the procedure described
above may lead to an underestimation of the distances (in a minority
of cases) and hence a bias in the inferred distribution of X-ray
luminosity. In this context, Fig. \ref{fig_5} shows the distribution
of \Lx for the ASBs divided into two subsets, 
namely those sources with or without Hipparcos parallax measurements.
For the sources with Hipparcos-derived  distances the measured $L_{X}$
span the range  10$^{28-32}$ \ergs, whereas for the remaining sources
the \Lx range extends to 10$^{32.5}$ \ergs (consistent with the fact
that the counterparts are typically somewhat fainter in the optical/NIR
and are potentially more distant).
The impact of an upward revision of \Lx  for the 10 sources where 
the dwarf versus giant classification is uncertain  (to the \Lx  
value corresponding to the quoted upper-bound on the distance) 
is also illustrated.  The main result is
an increase in the inferred number of systems with
$L_{\rm X} > 10^{32}$ \ergs~from 5 to 7. In practice, this
upperward shift in \Lx for a small number
of sources will have only a minor impact on the inferred XLF
of the sample of ASBs, taken as a whole (see \S3).


\begin{table*}
\small
\begin{center}
\caption{Details of the 46 sources in the current sample associated with ASBs.}
\begin{tabular}{cccccccccc}
\hline
XSS  & Distance & Distance & Type  & Name  &  Spectral Type & M$_K$ & Giant/ &  Hard Band &  log \Lx  \\
Name    &   pc     & Reference  &    &    &   & &  sub-Giant &  ct s$^{-1}$   &  erg s$^{-1}$ \\ 
\hline					   
J011635.6-023003 &  81 & (1) & RS CVn  & AY Ceti    & WD+G5IIIe    & -1.1  & Y & 1.4 & 30.84 \\
J012256.9+072509 &  45 & (1) & RS CVn  & AR Psc     & K1IV+G7V     & 1.7   & Y & 1.8 & 30.44 \\
J013500.9-295433 &  22 & (1) & BY Dra  & BB Sci     & K3-4V+K4-5V  & 3.0   & N & 0.6 & 29.32 \\
J023948.1-425302 &  36 & (1) & late-MS &   -        & M2V          & 5.6   & N & 0.8 & 29.90 \\
J024125.6+055916 &  42 & (1) & late-MS & BD+05378   & K6Ve         & 4.0   & N & 0.9 & 30.09 \\

J024325.4-375544 &  41 & (1) & BY Dra  & UX For     & G7V+K1V      & 2.9   & N & 1.3 & 30.20 \\
J044118.7+205403 &  13 & (1) & BY Dra & V834 Tau & K3Ve   & 4.5   & N & 0.6 & 28.91 \\
J045100.6-121416 &  300$_{-0}^{+80}$  & (5) & RS CVn? & HD30900 &  G7III & -0.2   & Y & 1.2 &  31.90 \\
J045330.4-555133 &  11 & (1) & late-MS & GJ2036A & M3Ve         & 6.1   & N  & 2.0 & 29.26 \\
J045818.2-751638 & 289 & (1) & FK Com  & YY Men   & K3IIIp       & -2.0  & Y & 1.0 & 31.80 \\

J050136.1-444949 &  94 & (1)  & FK Com & HD 32517    & G8/K0III     & 0.6   & Y & 0.7 & 30.68 \\
J050607.6-864153 &  64$_{-11}^{+17}$  & (2) & - & CPD-86 67  & KV      & 4.1  & N & 1.3 & 30.58 \\
J052845.4-652651 &  15 & (1) & late-MS  & AB Dor A   & K0V    & 3.8   & N & 2.4 & 29.62 \\
J054039.8-201759 & 221 & (1)  & RS CVn  & TW Lep     & F6IV+K2III   & -2.2  & Y & 1.1 & 31.61 \\
J062201.2-345448 & 250$_{-120}^{+200}$  & (5) & - &  -     & K       & 2.5   & ? & 0.6 & 31.43  \\

J063612.2+512258 & 80$_{-15}^{+20}$ & (5)     & late-MS? & -   & MV?      & 5.7   & N & 1.5 & 30.86 \\
J070846.2+554905 & 580$_{-250}^{+200}$ & (5)    & - & -         & K       & 2.4   & ?  & 0.6 & 32.20 \\
J081736.6-824336 & 14$_{-3}^{+4}$   & (2) & late-MS  & -        & M3.5V   & 5.9   & N & 0.6 & 28.95 \\
J082730.0-672401 & 750$_{-500}^{+30}$ & (5) & -  & -        & K      & 2.2   & ? & 0.6 & 32.42 \\
J092225.6+401200 &  25  & (1)  & BY Dra  & BF Lyn     & K3V+K3V     & 3.2   & N & 1.1 & 29.71 \\

J102146.8+605451 & 189 & (1)  & RS CVn  & FG UMa     & G9III        & -1.6  & Y & 2.1 & 31.75 \\
J130153.4-194631 & 121 & (1)  & Algol   & UY Vir     & A7V+G6IV     & 1.2   & Y  & 3.4 & 31.56 \\
J132132.4+385250 & 113 & (1)  & RS CVn  & BM CVn     & K2III        & -0.9  & Y & 0.5 & 30.71 \\
J133448.3+371058 &  46 & (1)  & RS CVn  & BH CVn     & F2IV+K2IV    & 0.3   & Y & 7.2 & 31.05 \\
J143851.3+194500 & 19$_{-4}^{+5}$ & (3)   & late-MS & - & M4V        & 7.6   & N  & 0.9 & 29.36 \\

J154547.5-302102 &  40  & (1) & BY Dra  & KW Lup     & K2V          & 3.4   & N & 1.6 & 30.27 \\
J155321.8-215817 & 172  & (1)  & Algol?   & HD14209    & A5V+?      & 1.0   & Y & 2.4 & 31.73 \\
J160900.7-190846 & 80$_{-15}^{+20}$ & (4)   & pre-MS?  &  & K9        & 4.6   & N  & 2.0 & 30.97 \\
J171725.8-665712 &  31  & (1) & RS CVn  & V824 Ara   & G7IV/V+K0IV/V & 2.2   & Y & 2.5 & 30.26 \\ 
J173240.2+741335 & 104  & (1)  & RS CVn  & DR Dra     & WD+K0III     & -1.0  & Y  & 1.1 & 30.94 \\

J175757.9+042732 & 600$_{-270}^{+80}$ &  (5) & -   &   -     & G     & 1.8   & ? &  0.9 & 32.38 \\
J182509.4+645018 & 159 & (1) & RS CVn? & HD170527   & K0 (III?)    & -1.6  & Y & 1.1 & 31.30 \\
J183355.8+514306 &  16 & (1)  & By Dra  & By Dra     & K6Ve+K7V     & 3.8   & N & 1.4 & 29.46  \\
J183419.9+184126 &  38 & (1)  & By Dra  & V889 Her   & G2V          & 2.9   & N & 1.1 & 30.07 \\ 
J184949.8-235011 & 3.0 & (1)  & late-MS  & V1216 Sgr & M3.5Ve       & 8.0   & N & 2.2 & 28.17 \\

J190825.4+522529 &  71 & (1)  & RS CVn  & V1762 Cyg  & K2IV-III+G8V & -1.3  & Y  & 2.5 & 30.97 \\
J195518.4+734933 & 325$_{-175}^{+235}$ & (5)   & - & -          & G/K    &  2.5  & ?  &  1.3 & 32.02 \\
J200030.9+592135 & 100$_{-20}^{+25}$   & (5)   & - & -          &  GV       & 3.3   & N  &  3.5  & 31.41 \\
J203654.3-465758 & 300$_{-230}^{+315}$  & (5)   & -    & -   &  K    & 1.8   & ? & 0.7   &  31.67 \\
J203722.8+705720 & 320$_{-120}^{+180}$  & (5)  & -       &    -      &  F  & 1.6   & ? & 0.9   & 31.82 \\

J204542.4+042141 & 120$_{-60}^{+100}$   & (5)  &  -      &   -       &  G  & 2.1 & ? &1.1   & 31.07 \\
J210225.8+274823 &  52 & (1)  & W UMa    & ER Vul     & G0V+G5V     & 2.1   & N  & 1.1 & 30.36 \\
J214126.5-104750 & 370$_{-290}^{+270}$ & (5)      & -  &   -       &  K & 1.6   & ? & 1.0  & 32.02 \\
J215627.2+051600 & 90$_{-40}^{+130}$ & (5)    & -  &   -       &  K  & 2.5  & ? & 0.5  & 30.50 \\
J231323.5+024033 &  97 & (1)   & RS CVn    & SZ Psc     & F8IV+K1IV    & -0.1   & Y & 3.0 & 31.32 \\

J234941.1+362531 &  130 & (1)  & FK COM  & OU And     & G1IIIe       & -1.5  & Y & 1.2 & 31.17 \\
\hline
\end{tabular}
\label{tab_1}
\end{center}
\raggedright{\bf References:}  {\bf(1)} Hipparcos parallax measurement - \citet{van07a};
{\bf(2)} \citet{riaz06};  {\bf(3)} \citet{lepine11}; {\bf(4)} Distance set at the lower limit 
of the range quoted by \citet{de99} for Upper Sco. Association ; {\bf(5)} this paper. \\
\end{table*}


\subsection{The cataclysmic variables} 

Table \ref{tab_2} provides summary information relating to the 16 probable
CVs in the current sample; the format is the same as in Table \ref{tab_1} 
except that the columns headed Spectral Type and Giant/sub-Giant are not included.
Roughly half of this sample (7 out of 16) are known magnetic CVs, \ie Polars (P)
or Intermediate Polars (IP), as opposed to the non-magnetic 
Nova-like Variables (NL) or Dwarf Novae (DN).

The source distance is again the key information required in the calculation of
the XLF.  Distance estimates and error bounds are available from the literature
(see the table for the references)  for 12 of the sources, 
including 5 based on the period-luminosity-colour prescription
employed by \citet{ak08}.  For the 4 sources 
lacking relevant information we have calculated the distance 
from the observed 2MASS $K_{s}$ magnitude assuming $M_{K} = 5.9 \pm1.3$,
representing the mean $M_{K}$ and standard deviation of the 
$M_{K}$ distribution of the 12 other sources.

In the colour-magnitude plane of Fig. \ref{fig_3}, the CVs in general
occupy the region above the locus of dwarf stars.  In Fig. \ref{fig_4}
the CVs are mostly located in same band as the ASBs, the
two exceptions being RBS 490 and TX Col. The anomalously-high transverse
velocity implied for RBS 490 has previously been noted by \citet{thor06}.

The last two columns of Table \ref{tab_2} give the measured X-ray count rate in 
the XSS hard band and the X-ray luminosity (log $L_{\rm X}$) determined from this
count rate and the assumed distance. Here we employ the calibration 
$1~\rm ct~s^{-1}$ = $7.9 \times 10^{-12} \rm~erg~s^{-1}~cm^{-2}$ 
(2--10 keV).  {\color{black} This is based on the assumption that the CV source
spectrum can be approximated as a $kT = 10$ keV thermal bremsstrahlung
spectrum subject to absorption in a line-of-sight column density of
$5 \times 10^{20}~\rm atom~cm^{-2}$; in fact, Fig. \ref{fig_2} suggests
that this spectrum is representative of the minimum
spectral hardness exhibited by the CV sample.}
The values of \Lx so determined span the range  10$^{30.5-33.5}$ \ergs~-
as illustrated in Fig. \ref{fig_5}.   Finally, the potential impact of the
uncertainties in the CV distances has been tested by revising all the distance
estimates to the upper bound quoted in Table \ref{tab_2} and recalculating
$L_{\rm X}$. As is evident from Fig. \ref{fig_5}, this results in a marginal upward
shift in the centroid of \Lx distribution.  However, in the context of the
CV XLF (discussed in \S3), this is not a significant effect.


\begin{table*}
\small
\begin{center}
\caption{Details of the 16 sources in the current sample associated with CVs}
\begin{tabular}{cccccccc}
\\
\hline
\hline
XSS  & Distance & Distance  &  Type  & Name  & M$_{K}$ &  Hard Band    &  log \Lx  \\
Name &   pc     & Reference &        &       &       &  ct s$^{-1}$  &  erg s$^{-1}$  \\ 
\hline					   

J014448.4+323258 & 147$_{-68}^{+128}$   & (1) & NL     & BG Tri    & 5.9 & 1.2 & 31.40 \\
J035410.3-165252 & 285$_{-105}^{+120}$ & (2,3)  & P & RBS 490  & 8.5 & 0.7 & 31.72 \\
J040910.7-711739 &  64$_{-17}^{+20}$  & (3) & DN  & VW Hyi & 7.7 & 1.1 & 30.63 \\
J053450.9-580136 & 848$_{-175}^{+219}$  & (4,5) & IP?  & TW Pic    & 4.5 & 2.7 & 33.26 \\
J054320.1-410201 & 585$_{-118}^{+155}$  & (4,6) & IP  & TX Col    & 4.3 & 1.5 & 32.69 \\

J061145.1-814925 & 158$_{-33}^{+42}$  & (4,7) & NL  & AH Men    & 5.9 & 1.0 & 31.36 \\
J062339.6-265750 & 190$_{-88}^{+165}$  & (1) & DN?      & -         & 5.9 & 1.4 & 31.68 \\
J062516.3+733439 & 573$_{-119}^{+146}$  & (4) & IP  & MU Cam    & 5.1 & 1.0 & 32.50 \\
J082513.7+730639 & 130$_{-24}^{+37}$  & (8) & DN  & Z Cam     & 5.3 & 1.2 & 31.27 \\
J082623.4-703143 &  90$_{-10}^{+45}$  & (9) & DN?  &      -    & 6.6 & 1.4 & 31.03 \\

J114336.8+714125 & 155$_{-35}^{+35}$  & (4,10) & IP  & DO Dra    & 6.4 & 1.4 & 31.52 \\
J130057.6-491211 & 105$_{-49}^{+91}$  & (1) & DN  & V1147 Cen & 5.9 & 1.6 & 31.23 \\
J173022.3-055935 & 359$_{-167}^{+313}$  & (1) & IP  & V2731 Oph & 5.9 & 1.3 & 32.19 \\
J185502.6-310937 & 510$_{-43}^{+52}$  & (11) & IP  & V1223 Sgr & 4.1 & 7.2 & 33.25 \\
J190715.9+440105 & 505$_{-50}^{+50}$  & (12) & NL  & MV Lyr    & 6.7 & 1.5 & 32.55 \\

J194904.4+774424 & 354$_{-73}^{+91}$  & (4) & DN  & AB Dra    & 5.4 & 1.0 & 32.08 \\

\hline
\hline
\end{tabular}
\label{tab_2}
\end{center}

\raggedright{\bf References:}  {\bf(1)} Assumes M$_K$ = 5.9 $\pm 1.3$, this paper; 
{\bf(2)}  \citet{thor06};
{\bf(3)}  \citet{pret12}; {\bf(4)} \citet{ak08}; {\bf(5)} \citet{norton00};
{\bf(6)}  \citet{buck89}; {\bf(7)}  \citet{gan99}; {\bf(8)} \citet{thor03};
{\bf(9)} \citet{parisi12}; {\bf(10)}  \citet{pret13}; {\bf(11)}  \citet{beu04};
{\bf(12)}  \citet{hoard04}. \\
\end{table*}


\section{2-10 keV X-ray Luminosity Functions}
\label{sec_3}

\begin{figure}
\centering
\includegraphics[width=6cm,angle=-90]{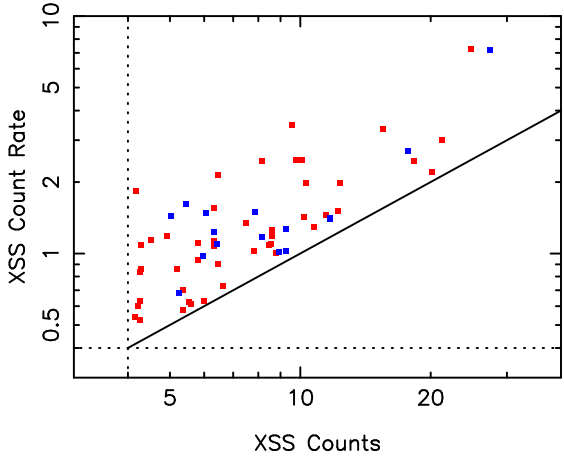}
\linespread{1}
\caption{The XSS hard-band count rate (in ct s$^{-1}$) plotted versus the
net counts (after background subtraction) recorded in the
XSS hard band.
The ASBs and CVs are represented by the red and blue points, respectively.
The vertical and horizontal dashed lines correspond to the criteria applied in
constructing the current sample, namely $\ge 4$ net counts and 
$\ge 0.4 \rm~ct~s^{-1}$. The solid diagonal line illustrates
the relationship between counts and count rate for a nominal 10 s exposure
time, which is close to the maximum in the XSS.  For all the sources in
the current sample, the net-count limit will always be encountered (as the
source distance is increased) before the minimum count-rate threshold is
reached. Hence the current sample can be described as count-limited. 
}
\label{fig_6}
\end{figure}

We have made use of the current sample to construct the differential XLF in the
2--10 keV band of the ASBs and CVs. As noted earlier, the
range of \Lx sampled spans roughly six decades from  10$^{28-34}$ \ergs.
For a flux-limited survey it is possible to estimate the space density of
sources in a given \Lx range using the 1/$V_{\rm max}$ method 
(\citealt{schmidt68}). In this approach, the distance $d_{\rm max}$ at
which a source would reach a given
flux-limit is calculated on the basis of the {\it measured} source
flux and  the known source distance.
The survey volume $V_{\rm max}$ is then determined by integrating
over the solid angle of the survey, taking into account
any variations in the flux-limit across the region.
However, in the present work it was
necessary to modify this approach  since the dominant selection
criterion is a {\it net counts} threshold rather than a specific 
count-rate threshold (the latter being akin to a flux limit)  
- as explained in Fig. \ref{fig_6}.
Our calculation of  $d_{\rm max}$, therefore, makes use of
the minimum count criterion of the survey, namely net 4 counts 
in the XSS hard band, along with the measured net counts in this
band registered for each source. 


A further consideration is that we are dealing with Galactic source
populations which  will show a spatial concentration towards the
Galactic plane.  
Assuming the space density of the sources declines as exp($-|z|/h$),
where $z$ is the distance above or below the Galactic plane
and $h$ is the population scale height, then the contribution to the volume
integral of a survey region of solid angle $\delta \Omega$ varies
with Galactic latitude $b$ as:

$
\\
\delta V_{\rm max} = \delta \Omega \frac{h^{3}}{{\rm sin}~|b|^{3}}[2 - (\xi^{2}+2\xi+2)e^{-\xi}] \\$

\noindent where $ \xi = d_{\rm max} \sin~|b|/h $  
(\citealt{tin93}; \citealt{saz06}; \citealt{pret12}; \citealt{pret13}).

In the present analysis we assume $h = 150$ pc for both the ASBs and CVs. 
\citet{saz06} make the same assumption and note that, for
their sample of ASBs, the inferred space density 
is not very sensitive to the scale-height estimate, since the majority of 
the sources are observed at $ z < 150$ pc (\ie less than 1 scale height).
For our current sample, 87 per cent of the ASBs
have inferred z-distances less than 150 pc, whereas for the
CVs, the fraction is 75 per cent. By way of comparison, 
\citet{pret12} and \citet{pret13} use scale height estimates of 120 pc
for long-period CVs and 260 pc for short period systems. 
We have investigated the impact of different scale height
assumptions in the present work; taking h = 120 pc (200 pc), instead of
150 pc, changes the inferred space density of 
the ASBs by a factor of $1.1$ ($0.9$), whereas for the CVs
the factor is $1.25$ ($0.75$).

We have calculated $V_{\rm max}$ for each source by summing the $\delta V_{\rm max}$
increments over the high latitude sky ($|b| > 10^{\circ}$), assuming the 
coverage fraction of the XSS is  35\% (\citealt{war12}). As a check
of the validity of this process we have also calculated $V/V_{\rm max}$ for 
each source (where $V$ is the volume obtained when the source distance 
$d$ is substituted for $d_{max}$ in the volume calculation). For the sample as a
whole, the distribution of $V/V_{\rm max}$ is reasonably uniform over the
range 0--1 with an average  $<V/V_{\rm max}> =0.514 \pm 0.031$, consistent
with the expectation for a homogeneous and uniform population.

\begin{figure}
\centering
\includegraphics[width=6.3cm,angle=-90]{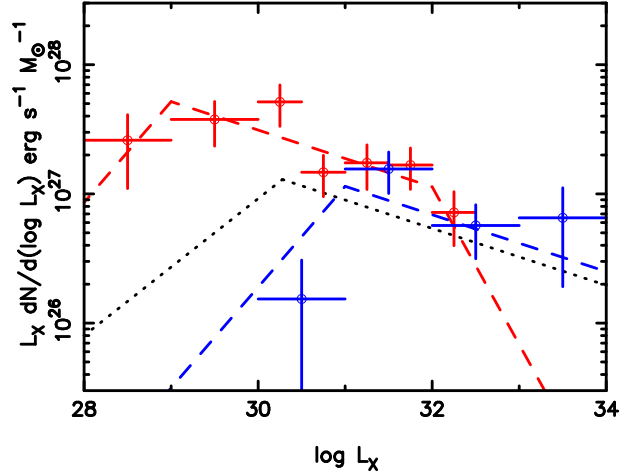}
\linespread{1}
\caption{The differential XLF in the 2--10 keV band for the ASBs
(red points) and the CVs (blue points), scaled by \Lx and normalized 
to the local stellar mass density (see text). The blue and red dashed lines
represent the best-fitting broken power-law fits to the 
data points.  The black dotted line shows the combined XLF for
ASBs and CVs derived by \citet{saz06}.
}
\label{fig_7}
\end{figure}

We have determined the source space density as $\sum (1/V_{\rm max})$, 
where the summation is over all the sources contributing to a particular
\Lx bin. In this context, although the total number of sources in
the current sample is relatively small, it has proved useful to derive
separate XLFs for the ASBs and CVs. The resulting
differential XLFs in the 2--10 keV band (employing logarithmic \Lx bins)
are shown as the two sets of points plotted in Fig. \ref{fig_7}. 
In this figure the y-axis has been scaling by the factor $L_{\rm X}/\rho_{0}$
where  $\rho_{0}$ is the stellar mass density in the solar neighbourhood.
Mirroring the approach used by \citet{saz06}, 
we set $\rho_{0} = 0.04 \rm~M_{\odot} \rm~pc^{-3}$. 
For the ASBs, we have utilized seven log \Lx bins encompassing the range  
10$^{28-32.5}$ \ergs, whereas for the CVs we use four such bins covering
the range 10$^{30-34}$ \ergs. 

We have fitted the central segment of each XLF with a power-law
function of the form:

$
\\
\frac{dN}{d\log L_{\rm X}} = K_{31} (L_{\rm X}/L_{31})^{-\alpha} \\ $

\noindent  where $\alpha$ is slope of the power-law and 
$K_{31}$ represents the normalization (scaled by $\rho_{0}^{-1}$) 
measured at an X-ray luminosity $L_{31} = 10^{31}$ \ergs. 
However, in order to match the
observed data, this power-law was modified, in the case of the ASBs,
by downward breaks (with $|\Delta \alpha| = 1$)
at both the low end and high end (more specifically at luminosities
of $10^{29}$ and $10^{32}$ \ergs, respectively).  The XLF of the CVs
was fitted similarly, but in this case with only a low-end
break at $L_{\rm X} = 10^{31}$ \ergs.
 
For the ASBs, trial fits of the doubly-broken power-law function
(varying $\alpha$ with $K_{31}$ as a free parameter) indicated a best-fitting
slope of $\alpha = 1.21 \pm 0.05$ (1$\sigma$ errors).  Since the error
range encompasses $\alpha = 1.22$, the value adopted by \citet{saz06},
as an aid to comparison  we have fixed $\alpha$ at this value for the 
ASBs. In the case of the CVs the error range on $\alpha$
was wider but again encompassed 1.22, so the XLF slope was also fixed
at this value for the CVs. 

The best-fitting normalizations of the XLF for the ASBs was
determined to be $K_{31} = 1.88 \times 10^{-4} $ M$_{\odot}^{-1}$ 
with a fitting uncertainty of roughly 15\%, whereas for the CVs the
value was $K_{31} = 1.15 \times 10^{-4} $ M$_{\odot}^{-1}$ to
a precision of roughly  25\%.  These are factors of $2.1 \pm 0.3$
and $1.3 \pm 0.3$ higher than the normalization
quoted by \citet{saz06}, for their combined XLF of ASBs and CVs.
The best-fitting broken power-law models of the XLFs
are shown in Fig. \ref{fig_7}, in comparison with
the XLF derived by \citet{saz06}.  

The inferred local space density for the ASBs obtained by integrating
over the full XLF is $3.4 \pm 0.5 \times 10^{-3}$ pc$^{-3}$. This is
higher than the space density reported by \citet{saz06}
due to both the upward shift in the XLF normalization and the extension
of the XLF power-law to a lower break luminosity.  The inferred local
space density of the CVs is $3.4 \pm 0.7 \times 10^{-5}$ pc$^{-3}$
which is toward the top end of the (albeit wide) range of
previous estimates (\eg \citealt{pret12} and references therein).

An upward trend is also evident in the integrated X-ray emissivities.
From the XLFs in Fig. \ref{fig_7} we obtain a 2-10 keV X-ray emissivity, 
scaled to the local stellar mass density, of $1.08 \pm 0.16 \times 10^{28}$ 
$\rm~erg~s^{-1}~M_{\odot}^{-1}$ for the ASBs and  
$2.5 \pm 0.6 \times 10^{27} \rm~erg~s^{-1}~M_{\odot}^{-1}$ 
for the CVs.  For comparison, \citet{saz06} quote a value of
$4.6 \pm 0.9 \times 10^{27} \rm~erg~s^{-1}~M_{\odot}^{-1}$  for ASBs 
(including young coronally-active stars) and CVs combined, whereas \citet{rev12}
determine a total 2-10 keV emissivity of 
$3.0 \pm 0.3 \times 10^{27} \rm~erg~s^{-1}~M_{\odot}^{-1}$ 
from an {\it RXTE} scan across the Galactic plane at 
$l =18.5^{\circ}$. 
The clear conclusion from the current analysis is that ASBs
make a more significant contribution to the hard-band source statistics
than previously recognised. 


\section{X-ray source counts on the Galactic plane}
\label{sec_4}

\subsection{Predicted Source Counts}

In this section we use our newly determined XLFs to investigate 
the contribution of ASBs and CVs to the 2--10 keV
X-ray source counts (\ie the log $N$ - log $S$ curve)  measured in the
Galactic plane. Specifically,
we focus on the direction ($l,b$) = ($28.5^{\circ}$,$0^{\circ}$) for
which the best observational constraints are available (outside of
the Galactic Centre region).

We assume that the current XLFs (which reflect the space density of sources
within 1 kpc of the Sun) apply across the whole of the Galactic disc, 
albeit with a normalization which scales with the local stellar mass density 
$\rho_{\rm gal}$. In the current work we estimate the latter using
a mass model for the Galactic disc of the form (\citealt{rev07}):

$
\\
\rho_{\rm gal}(R)~\propto~\exp [-(R/R_{\rm m})^{-3}-(R/R_{\rm disk})]
\\ 
$

\noindent where $R$ is Galactocentric radius, $R_{\rm disk}$ is the
exponential scale-length within the Galactic disc and $R_{\rm m}$
represents an inner cutoff to the disc, interior to which
the Galactic bulge and bar dominate. As noted by \citet{rev07},
for $l=28.5^{\circ}$ the contribution of the bulge/bar is likely
to be small and hence only the disc component need be considered.
We further note that the above expression does not include a 
dependence on $z$ and the $z$-scale height of the population, since
we restrict our source count calculation to a direction
on the Galactic plane. Following  \citet{rev07} we take $R_{\rm disk} = 2.2$ kpc, 
$R_{\rm m} = 2.5$ kpc and further assume that the stellar mass density
drops to zero for $ R > 10$ kpc. The scale factor applied to the
XLF at locations along the line of sight is then 
$\rho_{\rm gal}(R)/\rho_{\rm gal}(R_{sol})$, taking the distance 
to the Galactic centre, $R_{sol}$, to be 8 kpc.
As illustrated in Fig. \ref{fig_8}(a), for the specified direction, 
this scale factor reaches a maximum value
of $\approx 5$ at the tangent point (near 7 kpc) and
thereafter declines until the `disc edge' (at R = 10 kpc)
is encountered at a line-of-sight distance of 15.8 kpc.

Using the scaled XLFs, it is straightforward to calculate the {\it unabsorbed} 
flux distribution of the sources contained within a given
a volume element at a given distance. The conversion to observed
(\ie absorbed) flux then requires a model for the absorption along
the line of sight.  Here, for simplicity, we assume the X-ray
absorption arises from interstellar gas with a 
solar metal abundance and a density in the Galactic disc which
scales as the stellar-mass density. The gas density is normalized to a
solar-neighbourhood value of 0.4 atom cm$^{-3}$. The line of sight
through the Galactic plane  out to
15.8 kpc, then intercepts
a total column density of $5.8 \times 10^{22}\rm~atom~~cm^{-2}$,
consistent with earlier estimates (\citealt{ebi01}; \citealt{ebi05}).  
The reduction in the 2--10 keV flux due to absorption is plotted
as a function of the source distance in Fig.\ref{fig_8}(b).
Here, the transmission factors were calculate using 
WebPIMMS{\footnote{http://heasarc.gsfc.nasa.gov/Tools/w3pimms.html},
assuming the spectral forms for the ASBs 
and CVs detailed in \S2. The somewhat softer spectra
assumed for the ASBs compared to the CVs,
results in the lower transmission factors evident in
Fig.\ref{fig_8}(b). Our absorption model predicts a flux tranmission factor
of 0.57 for an ASB observed at the far edge of the Galactic disc,
compared to a value of 0.68 for a CV.


\begin{figure}
\centering
\includegraphics[width=4.5cm,angle=-90]{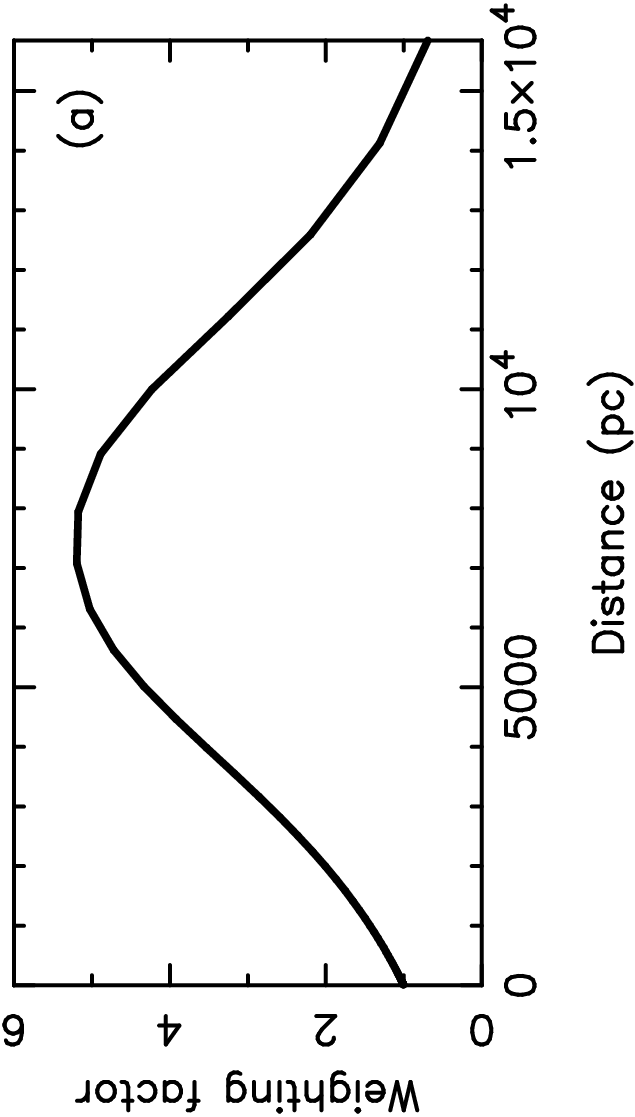}
\includegraphics[width=4.5cm,angle=-90]{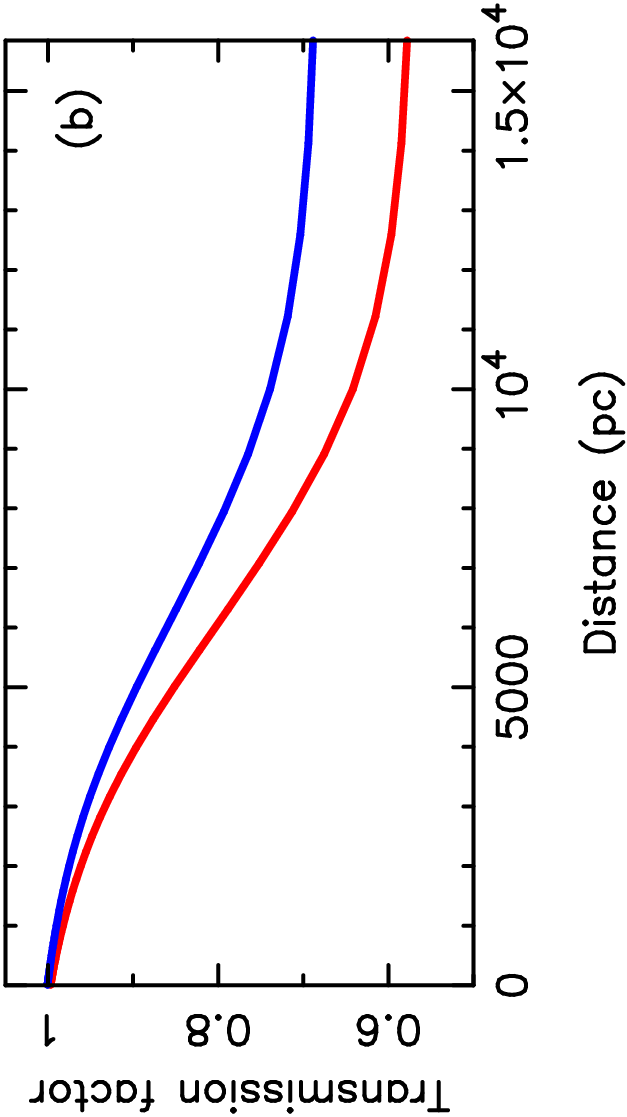}
\linespread{1}
\caption{ (a) The XLF scale factor as derived from
the model of the stellar mass density in the Galactic disk
plotted as a function of the line-of-sight distance in the
direction $(l,b)$ = ($28.5^{\circ}$,$0.0^{\circ}$). (b) 
The transmission versus distance for X-ray sources
in the Galactic plane for the same line of sight.
The values apply to fluxes measured in the 2--10 keV band.
The red curve corresponds to the ASBs
and the blue curve to the CVs.
}
\label{fig_8}
\end{figure}


Using the methodology outlined above, we have carried out an integration
along the line of sight to determine the source number versus flux 
relation for both the ASB and the CV populations.
However, to complete the picture we need
to included two further contributions, namely that of
Galactic XRBs and extragalactic X-ray sources.

For the XRBs we use a differential XLF similar to that discussed by
\citet{saz06} encompassing the range $ L_{\rm X} = 10^{34-38}$ \ergs.  
More specifically, we assume an XLF with $\alpha = 0.25$ up
to $ L_{\rm X} = 1.7 \times 10^{37}$ \ergs~ at which point the
slope steepens to  $\alpha = 0.9$. The normalization at the break point
was set at $2.6 \times 10^{-9} ~M_{\odot}^{-1}$.  The absorption versus
distance relation was taken to be the same as for the CVs.
The result was a good match to the high flux end of the observed
Galactic source counts as discussed in \S4.2.

For the extragalactic sources, we use the empirical formula
representing the source counts in the 2--10 keV band reported
by  \citet{mor03}. In this case we adjusted the flux scale by
a factor of 0.68 to account for the foreground absorption in the
Galactic plane.


\begin{figure}
\centering
\includegraphics[width=6.3cm,angle=-90]{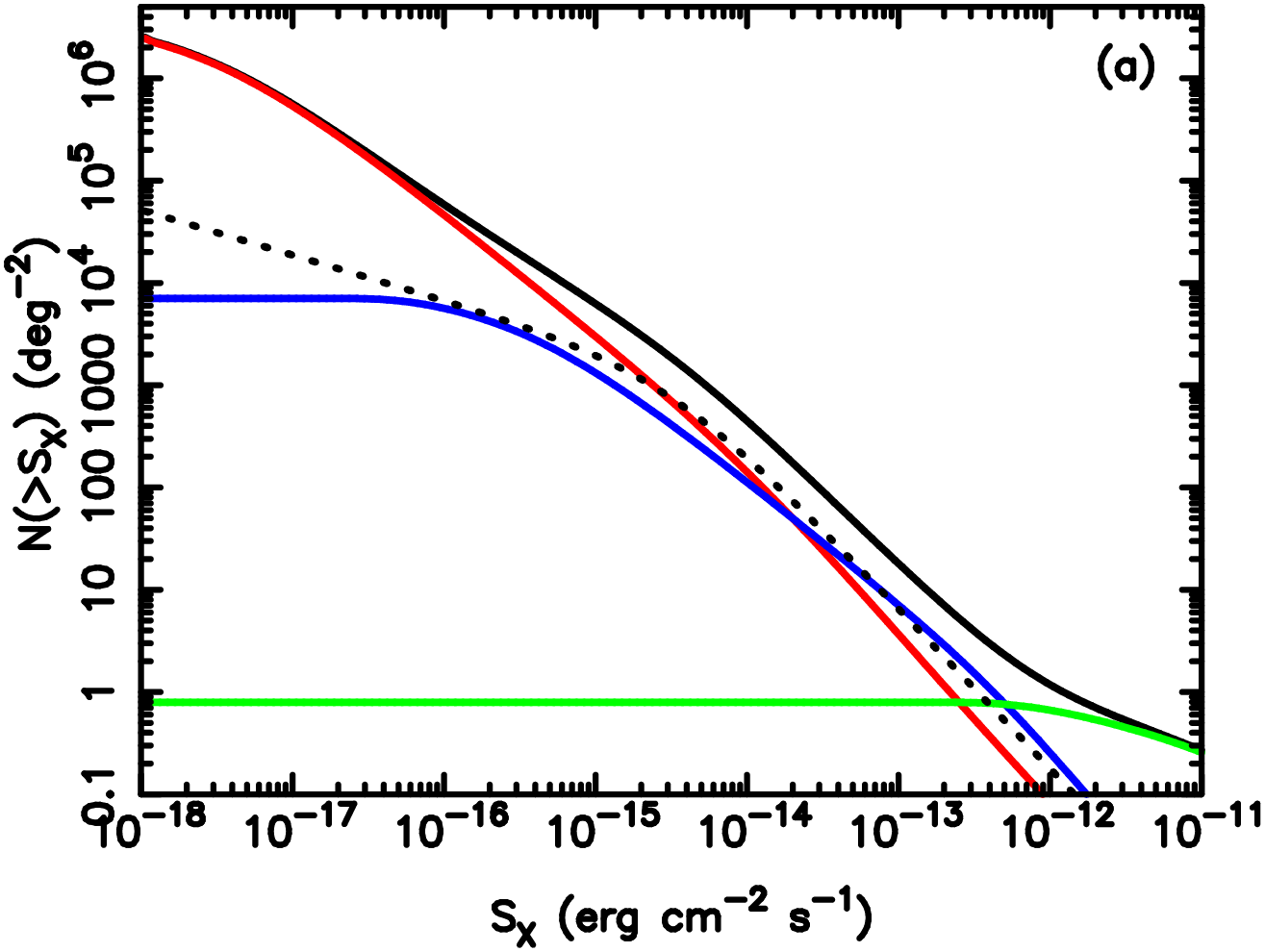}
\includegraphics[width=6.3cm,angle=-90]{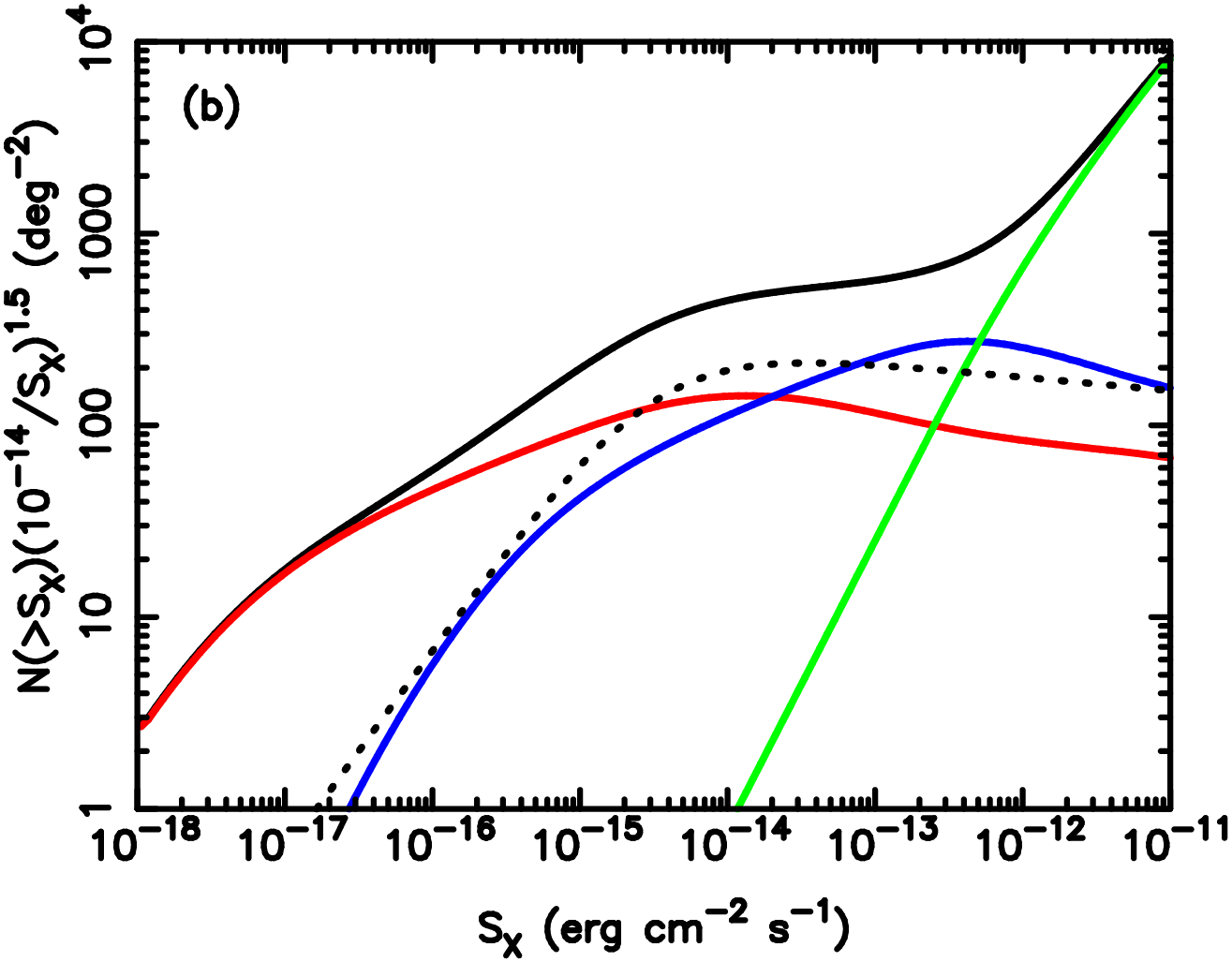}
\linespread{1}
\caption{(a) The predicted integral source counts in the
2--10 keV band. 
The number of sources observed down to a given 2--10 keV flux
threshold $S_{\rm X}$ is plotted versus  $S_{\rm X}$. The different curves
represent: the ASBs (red); the CVs (blue); the
XRBs (green); the extragalactic sources (dotted black) and
the combined counts (solid black).
(b) The predicted integral source counts
in 2--10 keV band normalized to the Euclidean form, \ie to the
function $(S_{\rm X}/10^{-14})^{-1.5}$.   
The curve representations are the same as in (a).
These predictions pertain to the direction 
($l,b$) = ($28.5^{\circ}$,$0.0^{\circ}$). 
}
\label{fig_9}
\end{figure}


The predicted number of sources is plotted versus the
limiting 2--10 keV flux ($S_{\rm X}$) in Fig. \ref{fig_9}(a).
As expected luminous XRBs dominate the integral
counts at the high-flux end. CVs become the numerically 
dominant population at intermediate fluxes, or more precisely in
the flux decade below $ S_{\rm X} = 5 \times 10^{-13}$ \ergscm.
Thereafter the integral counts are dominated by the extragalactic
sources until the ASBs come to the fore below
$ S_{\rm X} \approx  3 \times 10^{-15}$ \ergscm.  These crossover points
are further illustrated in the Fig. \ref{fig_9}(b),
which shows the corresponding log $N(>$$S_{\rm X})$ versus log $S_{\rm X}$
relation normalized to the function $(S_{\rm X}/10^{-14})^{-1.5}$ 
(\ie a source count relation with the {\it Euclidean} slope).
At the lower flux boundary in Fig. \ref{fig_9}(a), the predicted
number of ASBs per square degree is in excess
of $2 \times 10^{6}$.  At the far edge of the Galaxy, stars with
$S_{\rm X} = 10^{-18}$ \ergscm~have an intrinsic 2--10 keV X-ray
luminosity of $\approx 5 \times 10^{28}$ \ergs, implying that the
number counts of such systems will rapidly tail off
at even fainter fluxes (provided, of course, the inferred
low-end break in the XLF of the ASBs
is real).

\subsection{Comparison with Observations}

In recent times, X-ray observatories have surveyed many
segments of the low-latitude sky resulting in the detection of
large numbers of X-ray sources over a wide flux range
(\eg \citealt{muno03}; \citealt{muno06}; \citealt{rev07}; \citealt{rev11}; 
\citealt{mori13}; \citealt{nebot13}). Here we
focus on three observational datasets, which may be
reasonably compared with our source count predictions for the
direction ($l,b$) = ($28.5^{\circ}$,$0.0^{\circ}$).  The first 
stems from the extensive
survey encompassing the inner quadrant of the Galactic plane
carried out by ASCA (\citealt{sug01}).
The second relates to the {\it XMM-Newton} Galactic Plane Survey (XGPS),
which covered a narrow strip of the plane between
$ l = 19^{\circ}-22^{\circ}$ (\citealt{hands04}). Although
not specifically aligned with our target direction,
these surveys nevertheless provide very useful source detection
statistics in the flux range from $10^{-11}$ \ergscm~
down to roughly $3 \times 10^{-14}$ \ergscm (2--10 keV).
The limits on the Galactic log $N$ - log $S$ relation
resulting from these two surveys are shown in Fig. \ref{fig_10}.


\begin{figure}
\centering
\includegraphics[width=6.3cm,angle=-90]{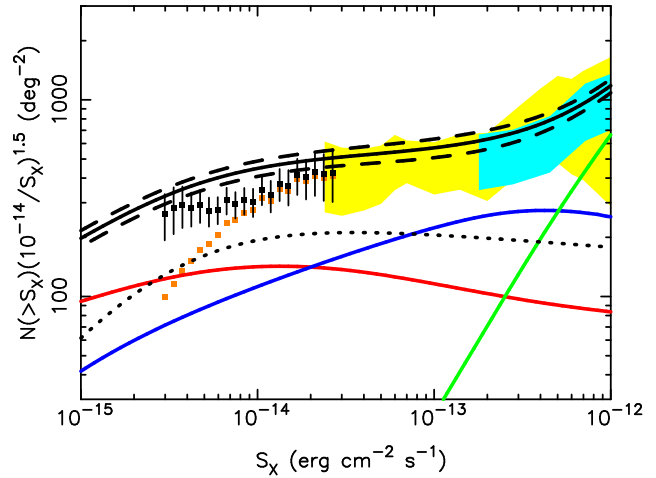}
\linespread{1}
\caption{The 2--10 keV Galactic log $N$ - log $S$ relation, normalize
to the function, $(S_{X}/10^{-14})^{-1.5}$. The light blue and yellow
shaded regions show the constraints derived from the ASCA and XGPS
surveys, respectively. The data points show the
number counts from two deep ($\sim 100$ ks) Chandra observations,
both before (brown squares) and after (black points with error bars)
applying a coverage correction - see text. The curves have the same
representation as in Fig. \ref{fig_9}(a). The combined (\ie total) counts 
(solid black curve) is shown with upper and lower error
bounds (dashed black curves).  
}
\label{fig_10}
\end{figure}


A pair of deep ($\sim 100$ ks), partly overlapping Chandra ACIS-I 
observations targeted at the plane near  $l = 28.5^{\circ}$ provide
important additional constraints. In these observations 
X-ray sources were detected with fluxes down to $\approx 3 \times 
10^{-15}$ \ergscm (2--10 keV), thus extending the source counts a decade
below that accessible via the XGPS programme.  \citet{ebi05} have
carried out a detailed study of the
these Chandra observations including the identification, where possible,
of NIR counterparts to the X-ray sources.
A interesting result from the \citet{ebi05} study (see also 
\citealt{ebi01}) was that the observed number density of X-ray sources,
at the survey sensitivity limit, is $\sim 600$ sources per square degree. 
These authors further note that this measured source density 
is comparable to that predicted for extragalactic sources,
after making due allowance for the impact of the
foreground absorption in the Galactic plane. Since this implied
rather tight constraints on the residual contribution of
Galactic sources at these faint fluxes, \citet{ebi05}
concluded that the bulk of the GRXE must be of
diffuse origin. Unfortunately, these conclusions are contrary to the
predictions based on our newly-derived XLFs for low-luminosity
Galactic X-ray sources.

To address this issue, we have re-assessed the spatial density of
the faint sources detected in the {\it Chandra} observations considered
by \citet{ebi05}. We take as the starting point the set of hard 
(3--8 keV) band detections reported by \citet{ebi05} (see their Table 1).
There are, in fact, 78 point sources detected in this band 
with a significance of $ \ge 4\sigma$. We convert the vignetting-corrected
hard-band count rates of these objects to a 2--10 keV flux assuming
the calibration 1 ct (100 ks)$^{-1}$ = $4 \times 10^{-16}$ \ergscm.
The application of a flux threshold of $3 \times 10^{-15}$ \ergscm (2--10 keV)
then results in the exclusion of 1 source from the sample. 
The implied sky density is 606 sources per square degree (taking the
sky area covered by the pair of overlapping fields to be 0.127 deg$^{2}$);
this (raw) estimate is fully consistent with the values reported
by \citet{ebi01} and \citet{ebi05}. However, this analysis neglects 
the corrections which must be applied for the varying
sensitivity across the Chandra field of view. 

We have estimated a coverage correction
(\eg  \citealt{muno03}; \citealt{bau04}; \citealt{mori13}) 
appropriate to the pair of {\it Chandra} observations using
the sensitivity map estimator provided in conjunction with the
Chandra Source Catalog \footnote{http://cxc.cfa.harvard.edu/csc/}.
Specifically, we used this tool to determine the
point source detection sensitivity (in units of $\rm photon~cm^{-2}~s^{-1}$)
in the {\it Chandra} H (2--7 keV) band for locations on a
$0.02^{\circ} \times 0.02^{\circ}$  square grid covering the combined
fields of view of the {\it Chandra} observations. These results were then
combined to produce a coverage curve (\ie the fraction of
the survey reaching a given sensitivity threshold),
against which the effective survey area for each source could
be determined (assuming the further calibration
1 ct (100 ks)$^{-1}$ = $4 \pm 0.4 \times 10^{-8} \rm~photon~cm^{-2}~s^{-1}$ in
the 2-7 keV band). 

The resulting source counts both before and
after apply this coverage correction
are shown in Fig. \ref{fig_10}. At the faint flux limit,
the number density of the {\it Chandra} sources rises to 
$1610 \pm 425$ deg$^{-2}$ 
after correction (with the relatively large
error bar reflecting uncertainties in the coverage correction
process).  This is well in excess
of the number density of extragalactic sources predicted
at  $S_{\rm X} = 3 \times 10^{-15}$ \ergscm  ($\sim 700$ deg$^{-2}$). 
The conclusion that the log $N$ - log $S$ curve of
Galactic X-ray sources continues to rise rapidly in
the faint flux regime, is also consistent with the number counts
reported by \citet{rev07}, albeit for sources
selected in the softer 1--7 keV band.

Fig. \ref{fig_10} shows the full set of constraints derived from the
observational datasets, in comparison with our estimates
for each class of source and the predicted total source count.
Upper and lower error bounds to the latter are also indicated;
these were calculated assuming a 15 per cent uncertainty on the
normalization of the ASB number counts, 25 per cent in case of the CVs,
and 10 per cent for both the XRBs and
extragalactic sources (all added in quadrature).
In general, the observed source number count and predicted total 
counts  are in good agreement. 
One caveat is, perhaps, that the Chandra data
show a small deficit,  relative to the total count prediction, at
$S_{\rm X} \approx 10^{-14}$ \ergscm. At this flux limit the
number of sources  actually {\it observed} in the pair of Chandra fields
is 37 compared to a {\it predicted} number of $48.0 \pm 4.3$;
the difference amounts to a $1.5\sigma$ effect.

\section{Contribution to the GRXE}
\label{sec_5}

Recent observations with {\it Suzaku} have added significantly
to our knowledge of the surface brightness distribution 
and spectral properties of the GRXE (\eg \citealt{yam09}; \citealt{uch13}).
In the context of the present work, the study carried out by \citet{ebi08}
is particular valuable in providing a precise measurement of the absolute
X-ray surface brightness in the direction ($l,b$) = 
($28.46^{\circ}$, $-0.20^{\circ}$).
\citet{ebi08} report that, after excluding point sources brighter than
$\approx 2 \times 10^{-13} \rm~erg~cm^{-2}~s^{-1}$ ~(2--10 keV), the total sky
brightness in the 2--10 keV band is 
$ 6.1 \times 10^{-11} \rm~erg~cm^{-2}~s^{-1}~deg^{-2}$.
After allowing for the contribution of the (extragalactic) cosmic X-ray
background, the resultant surface brightness of the GRXE in this band
is $4.8 \times 10^{-11} \rm~erg~cm^{-2}~s^{-1}~deg^{-2}$.

We have used the predictions from the previous section to estimate
the contribution that ASBs, CVs and XRBs make to the
GRXE.  Fig. \ref{fig_11} shows  the result in the form of the 'resolved'
fraction (of the GRXE) plotted versus the threshold flux. As is evident,
both from this figure and the source counts analysis (Fig. \ref{fig_9}),
the contribution of XRBs is negligible. Integrating down to an ultra-deep
flux threshold of  $10^{-18}$ \ergscm, the contribution
of the ASBs is $78 \pm 12$ per cent, with the CVs contributing
a further $16 \pm 4$ per cent. Clearly, a high fraction 
(and possibly the whole) of the  2--10 keV GRXE is resolved into
sources.

A further consideration is whether the integrated emission of ASBs
and CVs can explain the observed spectral form of the GRXE. We
will discuss this issue in detail in the next section. However, 
in the meantime we have repeated the above analysis for a harder spectral
range, namely the restricted 6-10 keV bandpass. 
Using the spectral forms previous defined for the two source
populations, we find that the fraction of the 2--10 keV flux contained
within this narrower band is 0.21 and 0.39 for the ASBs and CVs, respectively.
Similarly, using the information in \citet{ebi08}, we estimate the fraction
of the 2--10 keV GRXE surface brightness residing in the 6--10 keV band
to be 0.33. Applying these factors and assuming that the Galactic
absorption  in the 6--10 keV band is small, we find that the ASBs and CVs
contribute $ 62 \pm 10$ per cent and $ 21 \pm 5$ per cent of the
6--10 keV GRXE.   It follows that a very substantial fraction of
GRXE can be attributed to low-luminosity sources even in the hard
6--10 keV band.  We note, however, that this conclusion is (necessarily)
very dependent on the spectral form assumed for the ASBs, namely that 
of an {\bf apec} plasma with $T$ = 35 MK ($kT = 3$ keV)
and $Z = 0.4~Z_{\odot}$ (see \S2.1). Repeating the above calculations
for $kT = 1.5$ keV and $kT = 5$ keV changes the total ASB contribution
to the 6--10 keV GRXE to $13 \pm 2$ per cent in the case of the former,
to $90 \pm 13$ per cent for the latter. Thus, a factor 3 variation
in the assumed temperature takes us from under-prediction to
near over-prediction of the GRXE. 


\begin{figure}
\centering
\includegraphics[width=6.3cm,angle=-90]{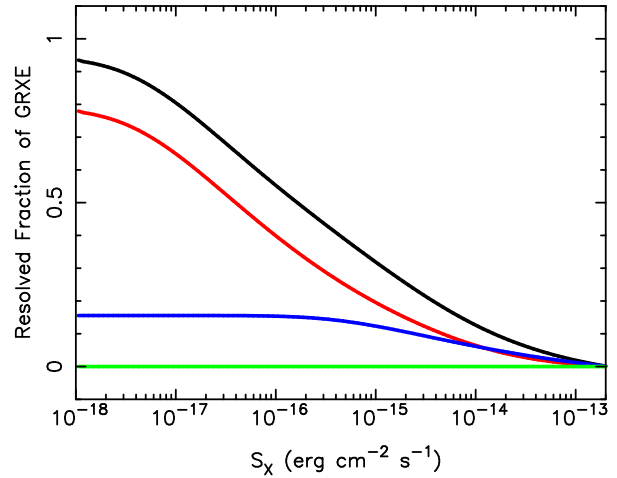}
\linespread{1}
\caption{The fraction of the 2--10 keV GRXE that is resolvable 
into point sources. The GRXE is assumed to have a surface brightness
of $4.8 \times 10^{-11} \rm~erg~cm^{-2}~s^{-1}~deg^{-2}$.
The contribution of the different source types is integrated from
a starting point at $2 \times 10^{-13}$ \ergscm~(2--10 keV) down
to faint fluxes.  The curves correspond to the ASBs (red), 
the CVs (blue), the XRBs (green) and the total (\ie combined)
contribution (black).
}
\label{fig_11}
\end{figure}


\section{Discussion}
\label{sec_6}

\subsection{Composition of the Galactic source population}

We have demonstrated that low-luminosity X-ray sources in
the form of ASBs and CVs dominate the hard-band source counts
recorded in the Galactic plane over a broad flux range. This is illustrated
in Fig. \ref{fig_12} which shows the {\it predicted} make-up of 
the X-ray source population detected in surveys on the Galactic plane
as a function of the survey depth. A first point to note
is that the contribution of extragalactic interlopers
is at its largest at $S_{\rm X} \sim 10^{-14}$ \ergscm, 
which is close to the median source flux in the 
{\it Chandra} observations discussed earlier.
The figure also illustrates how the composition of the Galactic population
changes as the survey depth increases.
Whilst the XRBs dominate at the bright end and
the ASBs come to the fore at the faint end, the 
CVs make their highest contribution, namely $\sim 40$ per cent of the
total source number, at intermediate fluxes
($S_{X} \sim 2 \times 10^{-13}$ \ergscm).
This latter point is consistent with recent results relating to
the X-ray source population at these
intermediate fluxes (\citealt{war14}).

\citet{ebi05} report the results from a NIR follow-up survey of their
{\it Chandra} observations carried out at ESO using the NTT with the
SofI NIR camera. This NIR survey
was quoted as being complete down to $\sim 18$, 17 and 16
mag in $J$, $H$ and $K_{s}$, which is roughly 2 magnitudes
fainter than 2MASS (\citealt{cutri03}; \citealt{skrutskie06}).  The mosaic
of NIR observations provided coverage of roughly two-thirds of the
field-of-view encompassed by the Chandra observations.
Using the results from \citet{ebi05} (their Table 1), we find that of
the 77 sources comprising the hard-band Chandra sample considered earlier,
57 were within the sky region covered by the new NIR survey.
Of these, 21 sources have plausible NIR counterparts, representing
a 37 per cent identification rate.  This  is very comparable
to the 0.32 fraction of X-ray detections predicted to be
ASBs at $S_{\rm X} \approx 10^{-14}$ \ergscm~(Fig. \ref{fig_12}). 

Of the 21 NIR counterparts, 16
have good photometry at $J$ and $K_{s}$ and hence can be
located on a NIR colour-magnitude diagram (see Fig. \ref{fig_13}).
This figure also shows the tracks of
several different types of stellar object in the colour-magnitude
plane assuming that the visual absorption, $A_{V}$, 
in the Galactic plane increases 
at the rate of 2 mag kpc$^{-1}$ (equivalent to $A_{J} = 0.56$ mag kpc$^{-1}$; 
$A_{K} = 0.224$ mag kpc$^{-1}$). It is evident that the NIR
parameters of the counterparts are consistent with those of
stars.  Interestingly 7 of these objects are aligned with the track
for a K3 giant, suggesting that these may be RS CVn systems
at the top-end of the XLF for the ASBs.  Two further
objects line-up with the track of a BOI supergiant and thus may be
sources in which the hard X-ray emission is generated in shocked regions
within an unstable wind or, in the case of massive binaries, 
in colliding winds (\citealt{mau10}; \citealt{war11}).
However, since the inferred 2--10 keV $L_{\rm X}$ 
is in excess of $10^{32}$ \ergs, the presence of an accreting high-mass
system cannot be excluded (\eg \citealt{and11}).


\begin{figure}
\centering
\includegraphics[width=6.3cm,angle=-90]{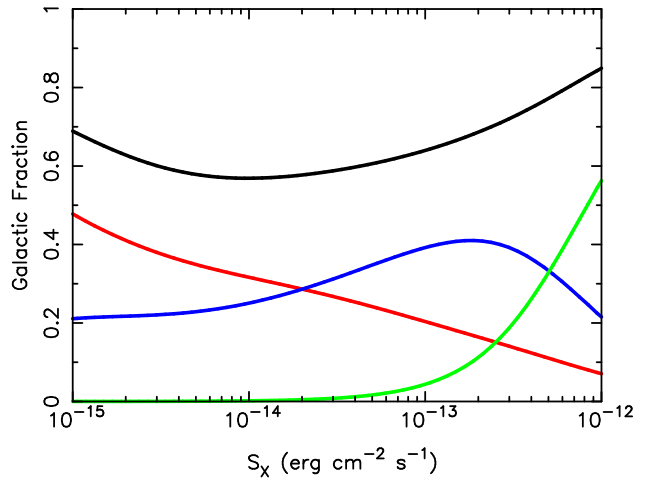}
\linespread{1}
\caption{The predicted composition of the X-ray source population observed
on the Galactic plane as a function of the limiting 2--10 keV flux.
The various curves show  the fraction of the population attributable to
different source types: ASBs (red curve);
CVs (blue curve); XRBs (green curve), all Galactic sources
(black curve).
}
\label{fig_12}
\end{figure}


\begin{figure}
\centering
\includegraphics[width=6.3cm,angle=-90]{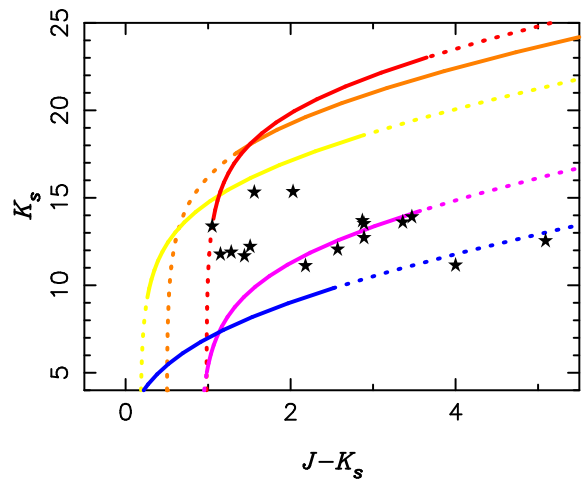}
\linespread{1}
\caption{The $K_{s}$ magnitude versus the $J-K_{s}$ colour 
of the likely counterparts of 16 hard-band 
{\it Chandra} sources taken from the 
survey of \citet{ebi05}
- shown as the black stars. 
The curves illustrate the tracks of different
types of star as the distance is varied from 1 pc - 20 kpc, assuming that
$A_{V}$ in the Galactic plane increases at the rate of 2 mag kpc$^{-1}$.
{\it Brown curve:} a luminous CV with M$_{K}$ = 5.0
and intrinsic $J-K_{s}$ = 0.5; {\it Yellow curve:} a FOV dwarf star with
M$_{K}$ = 2.25, $J-K_{s}$ = 0.18; {\it Red curve:} a
M6V dwarf with M$_{K}$ = 6.74,
$J-K_{s}$ = 0.98; {\it Purple curve:} a K3III giant with M$_{K}$ = -2.03,
$J-K_{s}$ = 0.90;  {\it Blue curve:} a BOI supergiant with 
M$_{K}$ = -6.44, $J-K_{s}$ = -0.15.
Apart from one exception, the curves become solid lines for
stellar distances in the range 300 pc - 9 kpc, which roughly
corresponds to an X-ray luminosity
in the range 10$^{29-32}$ erg s$^{-1}$, assuming an
X-ray flux of $10^{-14}$ \ergscm (2--10 keV). The exception is
the curve for the CV where the distance limits were set
at 3 and 16 kpc, corresponding to an X-ray luminosity in the
range 10$^{31-32.2}$.  This figure is adapted from
that shown in \citet{war14} (their Fig. 9).
}
\label{fig_13}
\end{figure}


Fig. \ref{fig_13} also illustrates that CVs will, typically, be
substantially fainter than the $K_{s} = 16$ magnitude limit of
the NIR survey. Similarly extragalactic sources are seen through
a total hydrogen column density of at least $6 \times 10^{22}$ cm$^{-2}$,
which equates to $A_{V} = 33$ using the standard relation
$N_{H}/A_{V} = 1.8 \times 10^{21}$~cm$^{-2}$~mag$^{-1}$ (\citealt{pred95})
and corresponds to $A_{J} \ge 9.2$ and $A_{K} \ge 3.7$. Hence the chances
of identifying extragalactic sources in the NIR in these {\it Chandra}
Galactic plane fields are remote. This substantiates the conclusion
reached above that the hard-band selected {\it Chandra} sources with
plausible NIR counterparts are predominantly ASBs.

\subsection{The origin of the hard GRXE}

In \S\ref{sec_5} we argued that ASBs
and CVS produce the bulk of the GRXE. The fraction of the GRXE accounted
for in our model amounted to $94 \pm 14$ per cent in the full 2--10 keV band,
falling to $83 \pm 11$  per cent in the restricted 6--10 keV band.
However, as demonstrated, these conclusions rely heavily on the validity
of the spectral assumptions underlying the calculations. Clearly, these
assumptions need to be justified.

We first consider the X-ray spectra of CVs.
In magnetic CVs, an accretion shock heats accreting material to high
temperature ($kT > 15$ keV). The resulting highly ionized plasma cools
in the post-shock flow and eventually settles on to the white dwarf surface 
via an accretion column. The resulting X-ray spectrum comprises a blend
of components produced at a range temperatures, densities and
optical depths with a 'characteristic' temperature 
typically in the range 10-20 keV (\eg \citealt{hel98};
\citealt{ezu99}; \citealt{cro99}; \citealt{yua10}).

Our (albeit small) sample of CVs is in fact comprised of
both magnetic and non-magnetic systems in roughly equal measure.
Non-magnetic CVs, which comprise the majority of 
the local CV population, are also well established sources of
X-ray emission at the lower end of the range of X-ray luminosity
exhibited by magnetic systems. In the current X-ray selected sample, 
the non-magnetic systems are, on average, roughly a factor
10 times less X-ray luminous than the confirmed magnetic systems.
In most non-magnetic CVs the X-ray flux is produced as thermal emission
from the boundary layer where the accretion flow slows down to
match the rotation of the white dwarf surface (\citealt{pat85}).
The X-ray emission will again be comprised of a blend of components 
with temperatures ranging from the shock temperature at the outer edge of
the boundary layer to the temperature of the white dwarf photosphere.
Observationally, the characteristic temperature of the X-ray continuum
emanating from non-magnetic CVs is typically in the range 5-- 20 keV.
(\citealt{pat85}; \citealt{bas05}; \citealt{ran06};
\citealt{byc10}; \citealt{reis13}).

In this paper we have modelled the X-ray spectra of the CVs
(both magnetic and non-magnetic) as a 10-keV thermal
bremsstrahlung continuum, which given the above,
would seem to be representative of this class of object.
 
We next consider the X-ray spectra of the ASBs. In active stars the
distribution of emission measure with temperature often shows a double
peak, with the hotter component often extending up to 30-40 MK, if not
higher (\citealt{gudel04}). Stellar X-ray surveys have also shown that there
is a relatively tight correlation between the characteristic coronal
temperature (as determined from low resolution spectral data)  of the 
`hotter' component and the normalized coronal luminosity $L_{\rm X}/L_{\rm bol}$ (\eg
{\citealt{gudel97}; \citealt{sch97}).
This correlation also appears to extend into the regime of stellar
flares, in the sense that the larger flares are generally hotter
(\eg \citealt{tel05}; \citealt{asc08}).
Large flares have durations typically ranging from hours to days with
total X-ray luminosities and temperatures reaching up to
$10^{33}$ \ergs~and 100 MK, respectively (\eg \citealt{pan97}; \citealt{tsu98};
\citealt{fav99}; \citealt{pan08}; \citealt{pan12}).
Clearly the occurrence of giant flares in
very active systems, for example RS CVn binaries, will enhance the 
detection rate of such sources in hard-band selected X-ray samples 
(\eg \citealt{pye83}; {\color{black}\citealt{matsuoka12}}). 
The fraction of the time that active systems 
spend in a flaring state is also relatively high. For example,
\citet{pan12} quote a flare-occurence rate
of $\sim 20$ per cent for RS CVn
binaries, whereas \citet{pan08} obtain a similar rate for flares on G-K 
dwarf stars. Finally, we note that recent high-resolution X-ray grating spectroscopy has
revealed that highly-active stars show the presence of an inverse 
`First Ionization Potential' effect, namely stellar spectra
with depleted abundances of elements such as Mg and Fe relative to 
solar photospheric values  (see \citealt{gudel04} - Fig 37).

Given the above properties of stellar coronae, our assumed spectral
model for the ASBs, namely a $T = 35$ MK ($kT =$~3 keV) {\bf apec} 
plasma with a metal abundance $Z = 0.4~Z_{\odot}$, would again seem
to be representative of this object class.

Although we have demonstrated that the integrated emission of unresolved
ASBs and CVs may well produce the bulk of the GRXE intensity in both 
the broad 2--10 keV band and the more restricted 6--10 keV range, 
a further consideration
is whether a mix of ASB and CV spectra can explain the observed Fe-line
features in the GRXE spectrum.  \citet{ebi08} measure
emission-line equivalent widths (EWs) for the He-like Fe K$\alpha$
line at 6.67 keV and the H-like Fe K$\alpha$ line at 6.96
keV of $350\pm40$ eV  and $70\pm30$ eV, respectively.
An Fe-K fluorescent line arising in neutral atoms or lightly
ionized ions was also observed near 6.4 keV with an
EW of $80\pm20$ eV.

For CVs, the line equivalent widths are found to be, typically, 
$\sim170$ eV for the 6.7-keV line and $\sim100$ eV for the
two other Fe lines, although with a very considerable source-to-source scatter
(\citealt{ezu99}; \citealt{hel98}; \citealt{hel04}; \citealt{ber12};
\citealt{war14}). In the case of the ASBs, the assumed 3-keV spectral model,
implies a He-like 6.7-keV Fe-K$\alpha$ line with 800 eV EW. However 
at this plasma temperature, there is negligible
H-like 6.9-keV Fe-line photon flux. In order to enhance the H-like
line, the temperature needs to rise by roughly a factor of two; for 
example for a  $Z = 0.4~Z_{\odot}$ plasma with $kT$ in the range 
5-7 keV, the EW of the 6.9-keV Fe-line line is between 90-140 eV.
The impact of such a temperature enhancement is illustrated in
Fig. \ref{fig_14}, which shows the 6.7-keV and 6.9-keV
Fe line EWs expected for a 3:1 mix of ASB and CV components, as the
temperature characterizing the ASB spectrum varies from 3 keV up to
7 keV.   It would seem entirely plausible that the ASBs present in
our XSS sample are in flare states characterized by this range of
temperature.


\begin{figure}
\centering
\includegraphics[width=6.3cm,angle=-90]{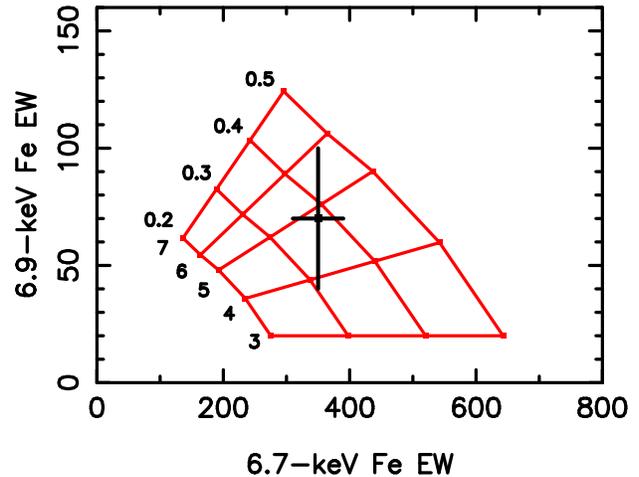}
\linespread{1}
\caption{The EW of the 6.9-keV H-like Fe K$\alpha$ line plotted
versus the EW of the 6.7-keV He-like Fe K$\alpha$ line. The
measurement (black point plus error bars) is from Suzaku
(\citealt{ebi08}). The grid (red lines) indicates the EWs
expected for a 3:1 mix of ASB and CV spectra on the basis of
different assumptions for the ASB spectra.  The grid points
correspond to coronal plasma temperatures ranging from 3 to 7 keV
in 1 keV steps with iron abundances ranging from 0.2 to 0.5 $Z_{\odot}$
in 0.1 $Z_{\odot}$ steps (as labelled). 
}
\label{fig_14}
\end{figure}


Although a suitable blend of hard X-ray emission from flaring ASBs together
with a more modest contribution from CVs has the potential to account for
the He-like and H-like Fe-K lines observed in the GRXE spectrum, the origin
of the Fe-K fluorescent line is more problematic. The 6.4-keV Fe fluorescent
line is very often too faint to be detected in the spectra 
of ASBs (\citealt{gudel09}), although some exceptions have been reported.
For example, a strong 6.4-keV line has been observed in several
protostars where the fluorescence is attributed to the irradiation
of the circumstellar disk by X-rays generated in hot coronal loops
(\eg \citealt{ima01}; \citealt{tsu05}; \citealt{ski07}).
There are also a limited number of reports of the detection of
6.4-keV emission in more evolved stars, where the fluorescence could
originate in the photosphere. For example, \citet{ost07} measured a prominent
6.4-keV Fe line in the spectrum of the RS CVn binary II Peg during a
superflare, with an EW ranging from 20 - 60 eV.
More recently, \citet{tes08} found evidence for photospheric
fluorescence in the single G-type star HR 9024 with a 6.4-keV line
of EW comparable to the top end value measured for II Peg.
However, notwithstanding these isolated observations, the hypothesis
that stellar flares might give rise
to the fluorescent emission present in the GRXE spectrum  (arising from Fe
and other elements) remains unsubstantiated. 

The 6.4-keV emission may in fact originate in an entirely different
way, namely in the scattering of the radiation of bright XRBs by the
intersteller medium. In a recent paper \citet{mol14} argue that between
10-30 per cent of the broad-band GRXE flux might originate in this way. The
results presented in this paper do not exclude this possibility, particularly
if the contribution from scattering in our fiducial direction
($l = 28.5^{\circ}$) is at the lower-end of the quoted range.
The scattered flux will be imprinted with a fluorescent Fe line
of high equivalent width ($\sim 1$ keV) for a scattering medium
with a near-solar Fe abundance (\citealt{sun98}). {\color{black} It follows that
even a modest $\sim10$ per cent contribution from scattered X-rays
to the continuum at 6 keV might explain the observed EW of the 6.4-keV
line feature seen in the GRXE spectrum.} 

{\color{black}
A further characteristic of the GRXE that merits consideration
is the observed fluctuation in its intensity on a scale of a few
degrees. For example, in the plot of the GRXE surface brightness
(4--10 keV) versus Galactic longitude presented by \citet{sug01} 
(their Fig. 2), peak-to-peak variations of up to 20 per cent
are evident on this angular scale  (measured with respect to a
smoothly varying Galactic profile).  This effect cannot
be explained in terms of statistical fluctuations in the number of
sources along the line of sight (within the context of our
Galactic-scale source distribution model), since the source number density 
at which a sizeable fraction of the GRXE intensity is produced 
is too high. It is more likely that these GRXE intensity
variations coincide with specific Galactic plane features
such as molecular-cloud complexes, OB associations and
star-formation regions. In this context our modelling of the Galactic
source population is most certainly incomplete, in that it excludes
the likely contribution to the GRXE of young stellar objects 
(e.g. \citealt{feig99}; \citealt{ima03}; \citealt{gudel04}) 
and massive stars (e.g. \citealt{mau10}; \citealt{and11}; \citealt{and14}),
neither of which were represented in the high-latitude XSS survey
which provides the basis of our study.

The picture which emerges is that ASBs and CVs likely give rise
to the bulk of the smoothly-varying GRXE (say nominally 
$\sim 80$ per cent of its typical intensity),
with a rough 3:1 split between the ASBs and CV contributions. Superimposed
on this, at the level of 10 percent or more, is the diffuse scattered X-ray emission
from bright XRBs, which carries the imprint of the 6.4 keV fluorescent line.
As a tracer of dense molecular gas and HI clouds, this component is predicted
to have a clumpy distribution  (\citealt{mol14}). Finally, a further
modest contribution (perhaps $\sim 10$ percent) may be contributed by other
relatively young Galactic source populations, which show a narrow
concentration on the plane in specific Galactic complexes and features.
Clearly the challenge for the future is to apply
fully quantitative tests of this description.
}

\section{Summary and conclusions}
\label{sec_7}

{\color{black} Using the XSS catalogue, we construct a sample of low-luminosity
Galactic X-ray sources.} This hard-band selected sample splits into two subsets,
namely 46 objects identified as coronally-active single stars and 
binaries (the ASBs) and a further 16 sources which
are likely CVs. We use published distances and/or our own distance estimates
to determine the distribution of X-ray luminosity within each source
class.  For the ASBs the 2--10 keV X-ray luminosities span the range
$10^{28-32.5}$ \ergs, whereas for the CVs, which include both magnetic
and non-magnetic systems in roughly equal numbers, the range is
$10^{30.5} - 10^{33.5}$ \ergs. 

We go on to determine the 2--10 keV X-ray Luminosity Function (XLF)
for each population and find that the XLF of the ASBs has a
normalization roughly a factor 2 higher than previous estimates and
extends at least a decade lower in X-ray luminosity
than previously projected. The 2-10 keV X-ray emissivity of the ASBs and CVs,
scaled to the local stellar mass density, is found to be 
$1.08 \pm 0.16 \times 10^{28}$ and  
$2.5 \pm 0.6 \times 10^{27} \rm~erg~s^{-1}~M_{\odot}^{-1}$ respectively. 
The implied total X-ray emissivity is at least a factor 2 higher than
previously reported.

We use the new XLFs to predict the 2--10 keV X-ray source
counts on the Galactic plane in the direction $ l = 28.5^{\circ}$,
a region for which a number of useful observational
datasets are available.  The predicted source counts are in good
agreement with the available observational constraints. As the flux
threshold is reduced from $10^{-12}$ \ergscm~down to $10^{-15}$ \ergscm,
XRBs, CVs, extragalactic sources and finally ASBs, each in turn, 
make the largest contribution to the
source number counts. 

{\color{black}
We also estimate the total X-ray surface brightness attributable to
the ASB and CVs source populations integrated along the line of
sight through the Galactic disc. We find that bulk the GRXE signal
($\sim 80$ per cent in the restricted 6-10 keV band) can be attributed
to the X-ray emission of these two populations, with a rough 3:1 split
between the ASBs and CV contributions. The predicted number density of
these sources is such that their integrated signal (after suitably
excluding the brightest sources) should be a smoothly
varying function of Galactic longitude (and latitude). Superimposed on
this will be the scattered X-ray emission from  bright Galactic XRBs
together with a (modest) contribution from relatively young
Galactic source populations. Together these  components might typical
contribute $\sim 20$ per cent of the total GRXE flux but with a clumpy
distribution tracing Galactic plane features and complexes.

Much ot the hard X-ray emission attributed to
the ASBs is likely to be produced during energetic flare states. 
Furthermore, it is reasonable to suppose that the hard-band selection
preferentially captures ASBs in such flaring episodes.
It seems plausible that the characteristic temperature of the
coronal loops which generate these flares will span the range 3-7 keV
and thereby generate the bulk of the He-like and H-like Fe-K$\alpha$
emission seen in the GRXE spectrum. However, the 6.4-keV Fe
fluorescent line seen in the GRXE spectrum is more likely to
originate in the diffuse scattered component referred to above.
}


\section*{Acknowledgements}

This paper is based on data from the \xmmn slew survey.
In carrying out this research, extensive use has been made
of ALADIN, VizieR and SIMBAD at the CDS, Strasbourg, France.
Similarly, the facilities provided by the Chandra X-ray Center
as part of the Chandra Data Archive proved very valuable.


\end{document}